\documentclass[aps,twocolumn,superscriptaddress,showpacs,showkeys,secnumarabic,citeautoscript,nofootinbib,nobibnotes,floatfix]{revtex4}

\usepackage[pdftex]{graphicx}
\usepackage{amsmath,amssymb,bm}
\usepackage{color}
\usepackage[usenames,dvipsnames]{xcolor}
\usepackage{hyperref}
\hypersetup{colorlinks=true,linkcolor=Red!90!Black,citecolor=Green!90!Black,urlcolor=Blue!90!Black,linktoc=page}
\usepackage{booktabs,tabularx}
\usepackage{float}

\usepackage[caption=false]{subfig}
\newcommand{\Tc}{T_{\rm\scriptscriptstyle C}}
\newcommand{\Hct}{H_{\rm{\scriptscriptstyle C}2}}
\newcommand{\FGL}{{\cal F}_{\rm\scriptscriptstyle GL}}
\newcommand{\JN}{{\bf J}_{\rm\scriptscriptstyle N}}
\newcommand{\JS}{{\bf J}_{\rm\scriptscriptstyle S}}
\newcommand{\tpsi}{\tilde\psi}
\newcommand{\tPsi}{\tilde\Psi}
\newcommand{\bpsi}{\bar\psi}
\newcommand{\tmu}{\tilde\mu}
\newcommand{\LV}[2]{U_{#1,#2}}
\newcommand{\LVU}[0]{\LV{x}{j}}
\newcommand{\LVV}[0]{\LV{z}{j}}
\newcommand{\LVW}[0]{\LV{y}{i}}
\newcommand{\LVZ}[0]{\LV{z}{i}}
\newcommand{\LVQ}[0]{\LV{x}{i}}

\begin{document}

\title{Stable large-scale solver for Ginzburg-Landau equations for superconductors}

\author{I.\,A.\,Sadovskyy}
\affiliation{
    Materials Science Division,
    Argonne National Laboratory,
    9700 S. Cass Avenue, Argonne, Illinois 60639, USA
}

\author{A.\,E.\,Koshelev}
\affiliation{
    Materials Science Division,
    Argonne National Laboratory,
    9700 S. Cass Avenue, Argonne, Illinois 60639, USA
}

\author{C.\,L.\,Phillips}
\affiliation{
    Mathematics and Computer Science Division,
    Argonne National Laboratory,
    9700 S. Cass Avenue, Argonne, Illinois 60639, USA
}

\author{D.\,A.\,Karpeyev}
\affiliation{
    Computation Institute,
    University of Chicago,
    5735 S. Ellis Avenue, Chicago, Illinois 60637, USA
}

\author{A.\,Glatz}
\affiliation{
    Materials Science Division,
    Argonne National Laboratory,
    9700 S. Cass Avenue, Argonne, Illinois 60639, USA
}
\affiliation{
    Department of Physics,
    Northern Illinois University,
    DeKalb, Illinois 60115, USA
}

\begin{abstract}
Understanding the interaction of vortices with inclusions in {type-II} superconductors is a major outstanding challenge both for fundamental science and energy applications. At application-relevant scales, the long-range interactions between a dense configuration of vortices and the dependence of their behavior on external parameters, such as temperature and an applied magnetic field, are all important to the net response of the superconductor. Capturing these features, in general, precludes analytical description of vortex dynamics and has also made numerical simulation prohibitively expensive. Here we report on a highly optimized iterative implicit solver for the time-dependent Ginzburg-Landau equations suitable for investigations of {type-II} superconductors on massively parallel architectures. Its main purpose is to study vortex dynamics in disordered or geometrically confined mesoscopic systems. In this work, we present the discretization and time integration scheme in detail for two types of boundary conditions. We describe the necessary conditions for a stable and physically accurate integration of the equations of motion. Using an inclusion pattern generator, we can simulate complex pinning landscapes and the effect of geometric confinement. We show that our algorithm, implemented on a GPU, can provide static and dynamic solutions of the Ginzburg-Landau equations for mesoscopically large systems over thousands of time steps in a matter of hours. Using our formulation, studying scientifically-relevant problems is a computationally reasonable task.
\end{abstract}



\date{\today}

\maketitle

\tableofcontents

\section{Introduction} \label{sec:introduction}

The time-dependent Ginzburg-Landau (TDGL) equations~\cite{Schmid:1966,Aranson:2002} are a powerful computational tool for describing the time-dependent dynamics of an order parameter near a phase transition that determine the macroscopic behavior of many system, most importantly {type-II} superconductors. They are an especially useful tool for addressing the important problem of designing superconducting materials that can support larger critical currents. As the energy dissipation in superconductors arises from the motion of vortices driven by the current-induced Lorentz force, this design problem requires developing a fundamental understanding of vortex dynamics. In contrast with computational models that treat the vortices as elastic strings moving in a viscous medium~\cite{Ertas:1996,Bustingorry:2007,Luo:2007,Koshelev:2011}, only the TDGL formulation can capture the correct interactions between pairs of vortices, vortices and inclusions, and allows vortices to cut and reconnect. The TDGL model provides a reasonable compromise between an approximate phenomenological and an exact microscopic descriptions of the vortex matter.

The strength of the TDGL formalism is that it describes the superconductor as a continuously distributed order parameter, given by a complex-valued scalar field. The amplitude of the order parameter is related to the local superconducting density in the system, which is suppressed by an array of singularities representing the vortices that arise spontaneously in the presence of a magnetic field. Pinning defects of arbitrary shape and size can be treated as either modulations of superconductor's critical temperature or by adding internal boundary conditions. The equations describing the evolution of the order parameter implicitly model the flexibility of the vortex lines, the long-range mutual vortex repulsion, vortex cutting and reconnecting, and even the interruption of current paths due to insulating pinning defects in the media. Although TDGL-based numerical simulations have been used many times to study properties of the vortex state~\cite{Doria:1990,Machida:1993,Crabtree:1996,Aranson:1996,Crabtree:2000,Winiecki:2002,Vodolazov:2013,Berdiyorov:2014}, a meaningful exploration of the phase space for realistic 3D superconductors has not been possible due to the computationally intensive nature of solving the equations at physically-relevant scales.

Here we address this challenge in two ways. First, we rewrite the TDGL equation in such a way that potential numerical instabilities are minimized and different boundary conditions can be implemented in a relatively simple way. In particular, the Poisson equation governing the scalar potential must be formulated in a stable and solvable way. Second, we implement an implicit iterative solver based on the Jacobi method. The result is a formulation of a solver for the TDGL equations where the value of the order parameter at each grid point of a mesh can be solved in a numerically stable manner using only information from the grid point's nearest neighbors. This type of formulation is critical for implementing a solver in a massively parallel computational environment. In contrast with other published work on solvers for the complex GL equations in massively parallel environments~\cite{Hawick:2011,Aleksic:2012}, we discuss all the aspects necessary to create a scientifically-relevant simulation, (e.g. magnetic fields, boundary condition, and current), each of which add non-trivial numerical complications to a system of equations. The increasing availability of massively parallel computational environments, especially in the form of hardware affordable to any researcher, (e.g. programmable general-purpose GPU devices) means that investigations of large-scale systems using the TDGL equations should no longer be computationally limited.

While  here we formulate our problem in the context of modeling a {type-II} superconductor, the TDGL equations are used to model a variety of problems, ranging from granular materials, self-propelled swimmers~\cite{Belkin:2010}, fluid and fracture dynamics~\cite{Dominguez:2013}, and cold atoms~\cite{Glatz:2011} to solidification from a melt or solution~\cite{Davis:2001}. Many aspects of the numerical formulation would also be relevant for solvers created for these problems as well. In particular, the presented methods are directly applicable to cold atom simulations that use complex coefficients in the Ginzburg-Landau (GL) equation, see e.g. Refs.~\cite{Glatz:2011,Scherpelz:2014}.

This paper is organized as follows. In Sec.~\ref{sec:equations_derivation} we review the TDGL equations and derive the formulation of the equations that we will use in subsequent sections. We simplify the equations by applying an $x$-axis aligned external current and assuming the large-$\lambda$ limit, which makes the magnetic field constant everywhere. In Secs.~\ref{sec:discretization} we introduce a discretization of the coupled partial differential equations and show they can be formulated as a set of linear equations that can be iteratively solved at each time step. In Secs.~\ref{sec:discretization_GL} and \ref{sec:discretization_Poisson} we present an implicit discretization schemes of the GL equations and the Poisson equation for the scalar potential, respectively. In Sec.~\ref{sec:boundary_conditions} we show how to implement periodic or open boundary conditions with respect to our discretization scheme. In Sec.~\ref{sec:pattern} we discuss how inclusions and different geometries can be modeled. In Sec.~\ref{sec:implementation} we discuss a computational implementation of this algorithm on a GPU in Sec.~\ref{sec:GPU} and provide several physically relevant examples of systems that can be investigated in Sec.~\ref{sec:results}. In Sec.~\ref{sec:conclusion} we make concluding remarks.

\section{Derivation of equations} \label{sec:equations_derivation}

\subsection{Ginzburg-Landau formalism} \label{sec:GL}

The Ginzburg-Landau (GL) equations describe spatial variations of the superconducting order parameter~$\psi$ in presence of electromagnetic vector potential ${\bf A}$. While a phenomenological theory, the GL theory can be rigorously connected to the microscopic  Bardeen-Cooper-Schrieffer theory in the vicinity of the critical temperature $\Tc$ of the superconducting phase transition~\cite{Gorkov:1959}. Far from $\Tc$, the GL equations do not correctly reproduce the physics in the vortex core, but still describe the interaction between vortices correctly, see e.g. Ref.~\cite{Aranson:2002}. Equilibrium states of superconductors are found from the GL equations by the minimization of the GL free energy, $\FGL$: ${\delta\FGL}/{\delta\psi^*}=0$ and ${\delta\FGL}/{\delta{\bf A}}=0$.

The TDGL equations are the dynamic generalization of the GL equations. For $\psi = \psi({\bf r},t)$ they can be written as
\begin{align}
    \Gamma \Bigl(\partial_t + \imath\frac{2e}{\hbar }\mu \Bigr) \psi
    & = a_0\epsilon ({\bf r})\psi -b|\psi |^2\psi
    \nonumber \\ & + \frac{1}{4m}\Bigl(\hbar \nabla
        + \frac{2e}{\imath c}{\bf A}\Bigr)^2\psi +\zeta ({\bf r},t),
    \label{eq:GL} \\
    \nabla \times (\nabla \times {\bf A})
    & = \frac{4\pi}{c} \bigl[ \JN
        + \JS + {\cal I}({\bf r},t) \bigr],
    \label{eq:Maxwell}
\end{align}
where $-e$ and $m$ are the electron's charge and mass, $c$ is the speed of light, $\mu$ and ${\bf A}$ are the scalar and vector potentials, $\Gamma$, $a_0$, $b$ are phenomenological constants that can be derived from the microscopic theory, and $\imath$ is the imaginary unit. The Langevin terms $\zeta ({\bf r},t) $ and ${\cal I}({\bf r},t)$ describing thermal noise have the correlators
\begin{align}
    & \langle\zeta^*({\bf r},t) \zeta({\bf r}',t') \rangle
    = 2 \Gamma T \, \delta({\bf r} - {\bf r}' ) \delta(t - t'),
    \label{eq:noise_zeta} \\
    & \langle {\cal I}_\alpha({\bf r},t) \, {\cal I}_\beta({\bf r}',t')\rangle
    = 2T\sigma\delta_{\alpha\beta} \, \delta({\bf r} - {\bf r}') \delta(t - t'),
    \label{eq:noise_I}
\end{align}
respectively, where~$\sigma$ is the normal conductivity.

Equation~(\ref{eq:Maxwell}) is the Maxwell equation\footnote{Note, that the term $(1/c) \, \partial_t {\bf E}$ in Eq.~(\ref{eq:Maxwell}) is neglected because the time variation rates of ${\bf A}$ are much slower than the frequencies of the corresponding electromagnetic waves.} governing the dynamics of the~${\bf A}$, where $\JN$ and $\JS$ are normal and superconducting currents, respectively. These currents are given by
\begin{align}
    \JN
    & = -\sigma \bigl[ ({1}/{c}) \, \partial_t{\bf A} + \nabla \mu \bigr],
    \label{eq:Jn} \\
    \JS
    & = -\frac{e}{2m}\Bigl[ \psi^*\Bigl(
        \imath\hbar \nabla + \frac{2e}{c} {\bf A}
    \Bigr) \psi + {\rm c.c.}\Bigr].
    \label{eq:Js}
\end{align}

We use the Landau gauge fixing condition $\nabla {\bf A} = 0$. Correspondingly, the magnetic induction ${\bf B}$ and electric field ${\bf E}$ are determined by ${\bf B} = \nabla \times {\bf A}$ and ${\bf E} = -(1/c) \, \partial_t{\bf A} - \nabla\mu$.

A spatially dependent critical temperature~$\Tc({\bf r})$ can be used to model large-scale inhomogeneities and pinning sites. In Eq.~(\ref{eq:GL}), $\epsilon({\bf r})$ is a dimensionless function defined as
\begin{equation}
    \epsilon({\bf r})
    = [\Tc( {\bf r}) - T] / T
    \label{eq:epsilon}
\end{equation}
and which therefore vanishes at the local critical temperature $T \to \Tc({\bf r})$. This modeling is covered in detail in Sec.~\ref{sec:pattern}. Alternatively, pinning sites can be modeled by introducing voids of various shapes in the integration domain and imposing appropriate internal boundary conditions.

At zero temperature, the GL theory of superconductivity involves two characteristic length scales: the coherence length
\begin{equation}
    \xi_0 = \sqrt{\hbar^2 / 4 m a_0},
    \label{eq:coherence_length}
\end{equation}
and the magnetic penetration length
\begin{equation}
    \lambda_0 = \sqrt{m c^2 / 8\pi e^2 \psi_0^2 },
    \label{eq:penetration_depth}
\end{equation}
where $\psi_0= \sqrt{a_0/b}$ is the equilibrium value of the order parameter in the absence of an electromagnetic field. For a spatially uniform $\Tc({\bf r}) = \Tc$ and, therefore, constant $\epsilon({\bf r}) = \epsilon$, the temperature-dependent coherence length $\xi = \xi_0 / (1 - T/\Tc)^{1/2}$ describes the typical scale of order parameter variations in space and the magnetic penetration length $\lambda = \lambda_0 / (1 - T/\Tc)^{1/2}$ describes the depth to which a small external magnetic field can penetrate the superconductor. The temperature-independent ratio $\varkappa = \lambda / \xi = \lambda_0 / \xi_0$ is called the GL parameter.

Strictly speaking, these dynamic equations are microscopically justified only for gapless superconductors~\cite{Gorkov:1968}. However, the TDGL description of the dynamics can also be used to study static or slowly moving vortex configurations (steady states).

For numerical simulations, Eqs.~(\ref{eq:GL}) and (\ref{eq:Maxwell}) can be written in the dimensionless form
\begin{align}
   & u(\partial_t + \imath\mu)\psi
     = \epsilon ({\bf r})\psi - |\psi |^2\psi
        + (\nabla - \imath{\bf A})^2\psi + \zeta ({\bf r},t),
    \label{eq:GL_dimensionless} \\
    &\varkappa^2\nabla \times (\nabla \times {\bf A})
     = \JN
        + \JS + {\cal I}({\bf r},t),
    \label{eq:Maxwell_dimensionless}
\end{align}
where $u = \Gamma / a_0 t_0$ with the unit of time $t_0 = 4\pi \sigma\lambda_0^2/c^2$. The zero-temperature coherence length $\xi_0$ is used for the unit of length. Correspondingly, the total current density ${\bf J} = \JS + \JN$ (in units of $J_0= \hbar c^2 / 8\pi e \lambda_0^2 \xi_0 = (e\hbar/m\xi_0)\psi_0^2$) assumes the form\footnote{The maximum supercurrent density which can flow without dissipation (depairing current density) is $J_{\rm dp} = (2 / 3\sqrt{3}) \, (1 - T/\Tc)^{3/2} \approx 0.385 \,(1 - T/\Tc)^{3/2}$ in the reduced units.}
\begin{equation}
    {\bf J} = {\rm Im} [ \psi^*(\nabla - \imath{\bf A})\psi ]
        - (\nabla \mu + \partial_t {\bf A}),
    \label{eq:J}
\end{equation}
where the magnetic field is measured in units of the upper critical field $\Hct(0) = \hbar c / 2e \xi_0^2$ at zero temperature and the electric field  ${\bf E} = - \nabla \mu - \partial_t {\bf A}$ is measured in units of $\xi_0 \Hct(0) / (c t_0)$. The correlation properties of the reduced noise terms will be discussed in detail below.

In many superconductors, most prominently in cuprate high-temperature superconductors, the coherence length in the $ab$ plane is different than in $c$ direction. This anisotropy can be modeled in Eqs.~(\ref{eq:GL})--(\ref{eq:Js}) by treating the mass and conductivity as anisotropic parameters, e.g. using an ``effective mass'' $m \to {\rm diag} \{1, \, 1, \, g^2 \} m$ and ``effective conductivity'' $\sigma \to {\rm diag} \{1, \, 1, \, 1/g^2 \} \sigma$, where $g$ is an anisotropy parameter. In this case, the component of the Laplacian in Eq.~(\ref{eq:GL_dimensionless}) and gradients in Eq.~(\ref{eq:J}) should be changed
to
\begin{align*}
    &(\nabla - \imath{\bf A})^2  \to
    (\partial_x - \imath A_x)^2 + (\partial_y - \imath A_y)^2 + (\partial_z - \imath A_z)^2/g^2,
    \\
    &(\nabla - \imath{\bf A}) \to
    \bigl[ \partial_x - \imath A_x, \; \partial_y - \imath A_y, \; (\partial_z - \imath A_z)/g^2 \bigr]^{\tt T},
    \\
    &\nabla\mu \to
    \bigl[ \partial_x \mu, \; \partial_y \mu, \; \partial_z \mu / g^2 \bigr]^{\tt T}.
\end{align*}
Additionally, the $z$-component,  ${\cal I}_z$, of the fluctuation current ${\cal I}$ in Eq.~(\ref{eq:Maxwell_dimensionless}) is rescaled to ${\cal I}_z/g$.

\subsection{External current} \label{sec:external_current}

In general, we are interested in determining the current-voltage characteristic of the superconductor as a function of different conditions. We want to apply an external current to the system, which will produce a voltage drop or electric field across the superconductor. Without loss of generality, we choose a current in the $x$ direction, resulting in the average electric field having a non-zero $x$ component, $E_x$, and apply periodic boundary conditions in $x$ direction.

Under these conditions, the scalar potential $\mu$ is, on average, an increasing or decreasing function of $x$ depending on the sign of the applied current, creating a discontinuity at the boundary in the $x$ direction. This is resolved by removing the zero mode from $\mu$ and adding a phase factor to the order parameter. This leads to the following gauge transformations:
\begin{align}
    \mu({\bf r}) & = - x E_x + \tmu({\bf r}),
    \label{eq:mu_transformation} \\
    \psi({\bf r}) & = \tpsi({\bf r}) \, e^{\imath Kx},
    \label{eq:psi_transformation}
\end{align}
where $\tmu$ a is periodic function, $\tpsi$ is a quasi-periodic function (see Sec.~\ref{sec:bc_periodic} for details), and $K = K(t)$ does not depend on~$\bf r$. Using Eqs.~(\ref{eq:mu_transformation}), (\ref{eq:psi_transformation}), and (\ref{eq:GL_dimensionless}), we obtain
\begin{align*}
    & u[\partial_t + \imath(\partial_t K - E_x) x
    + \imath\tmu] \tpsi =
    \nonumber \\ & \qquad\qquad
    \epsilon({\bf r})\tpsi
    - |\tpsi|^2\tpsi
    + (\nabla + \imath{\bf K}
    - \imath{\bf A})^2\tpsi + \zeta({\bf r},t),
\end{align*}
where ${\bf K} = [K, \, 0, \, 0]^{\tt T}$. By choosing $\partial_t{K}$ to be the electric field, or
\begin{equation}
    \partial_t{K} = E_x \label{fig:KttoE},
\end{equation}
Eqs.~(\ref{eq:GL_dimensionless}) and (\ref{eq:J}) become
\begin{align}
    u(\partial_t + \imath\tmu)\tpsi
    & = \epsilon({\bf r})\tpsi
    - |\tpsi|^2\tpsi
    + (\nabla - \imath\tilde{\bf A} )^2 \tpsi
    + \zeta({\bf r},t),
    \label{eq:GL_external_current} \\
    {\bf J} & = {\rm Im}[\tpsi^*(\nabla - \imath\tilde{\bf A})\tpsi]
    - \partial_t \tilde{\bf A} - \nabla \tmu,
    \label{eq:J_external_current}
\end{align}
where  $\tilde{\bf A} = {\bf A} - {\bf K}$ is a generalized vector potential.

If electroneutrality is assumed, then the total current has to be conserved, $\nabla {\bf J}=0$. This condition and Eq.~(\ref{eq:J_external_current}) together with the Landau gauge lead to the Poisson equation for~$\tmu$,
\begin{equation}
    \Delta\tmu =
    {\bf\nabla} \,
    {\rm Im} [ \tpsi^*({\bf\nabla}-\imath\tilde{\bf A})\tpsi ].
    \label{eq:Poisson}
\end{equation}
We can now use Eq.~(\ref{eq:J_external_current}) to define a differential equation for $K$. The $x$-component of the current can be explicitly written as
\begin{equation}
    J_x
    = {\rm Im} [ \tpsi^*
        (\partial_x + \imath K - \imath A_x)\tpsi ]
    + \partial_t{K}-\partial_t A_x
    - \partial_x\tmu.
\end{equation}
This expression averaged over any $yz$ cross section (or the whole system) then must equal to the external applied current,
\begin{align}
    J_{x,\rm ext}
    = \langle J_x \rangle
    & = {\rm Im} [ \langle \tpsi^*
        (\partial_x + \imath K - \imath A_x)\tpsi \rangle ]
    + \partial_t K-\langle \partial_t A_x \rangle.
    \label{eq:J_external}
\end{align}
Here we took into account that $\langle \partial_x \tmu \rangle=0$ due to the periodic boundary conditions. This gives a differential equation for $K$, which we rewrite as
\begin{equation}
    \partial_t K-\langle \partial_t A_x \rangle + J_{\rm av} - J_{x,\rm ext} = 0,
    \label{eq:K_ODE}
\end{equation}
where $J_{\rm av}$ is the supercurrent averaged over whole system
\begin{equation}
    J_{\rm av} = {\rm Im} [ \langle \tpsi^*(\partial_x
        + \imath K - \imath A_x)\tpsi\rangle ].
    \label{eq:J_av}
\end{equation}

Equations~(\ref{eq:K_ODE}) and (\ref{fig:KttoE}), therefore, define the electric field response of the system to the applied current $J_{x,\rm ext}$. Since $\tpsi$ depends implicitly on $K$, Eq.~(\ref{eq:K_ODE}) cannot be solved analytically and needs to be numerically integrated instead.

In dissipative states, where $J_{\rm av}$  is not equal $J_{x,\rm ext}$,  $K(t)$ increases with time and causes large oscillations of the real and imaginary parts of the order parameter, resulting in numerical instabilities. However, on a discrete regular grid, when the oscillations are on the order of the grid spacing $h_x$, i.e., $Kh_x \geqslant 2\pi$, one can {\it rewind} the phase of the order parameter by replacing $K \to K - (2\pi/h_x) \lfloor Kh_x/2\pi\rfloor$, where $\lfloor\cdot\rfloor$ stands for the floor function. Thus, as the equations are integrated forward in time, $K$ is never permitted to get too large.

\subsection{Simplifications in the large-$\lambda$ limit} \label{sec:large_lambda}

For high-$\Tc$ superconductors or thin superconducting films, the penetration length is much larger than the coherence length, $\lambda \gg \xi$. In this case, when the distance between vortex lines is smaller that $\lambda$, the magnetic field ${\bf B}$ can be considered homogeneous.

In our formulation of the problem, we will consider two cases for the magnetic field orientation: (i) in the $xz$ plane
\begin{subequations}
\begin{equation}
    {\bf B}^{xz} = [B_x, \; 0, \; B_z]^{\tt T}
    \label{eq:B_xz}
\end{equation}
or (ii) in the $yz$ plane
\begin{equation}
    {\bf B}^{yz} = [0, \; B_y, \; B_z]^{\tt T}.
    \label{eq:B_yz}
\end{equation}
\end{subequations}

Without significantly restricting the problem space, assuming either the~$x$ or $y$ component of the magnetic field is zero simplifies the vector potential. Now dependent on only one spatial coordinate, the vector potentials are given by
\begin{subequations}
\begin{align}
    {\bf A}^{xz} & = \bar y \, [-B_z, \; 0, \; B_x]^{\tt T},
    \label{eq:A_xz} \\
    {\bf A}^{yz} & = \bar x \, [0, \; B_z, \; -B_y]^{\tt T},
    \label{eq:A_yz}
\end{align}
\end{subequations}
where $\bar x = x - L_x/2$ and $\bar y = y - L_y/2$, $L_x$ and $L_y$ are system sizes in $x$ and $y$ directions, respectively.  In particular $\partial_t {\bf A}=0$ henceforth, simplifying Eq.~(\ref{eq:K_ODE}).

\begin{figure*}[tb]
    \begin{center}
        \subfloat{\includegraphics[width=14.0cm]{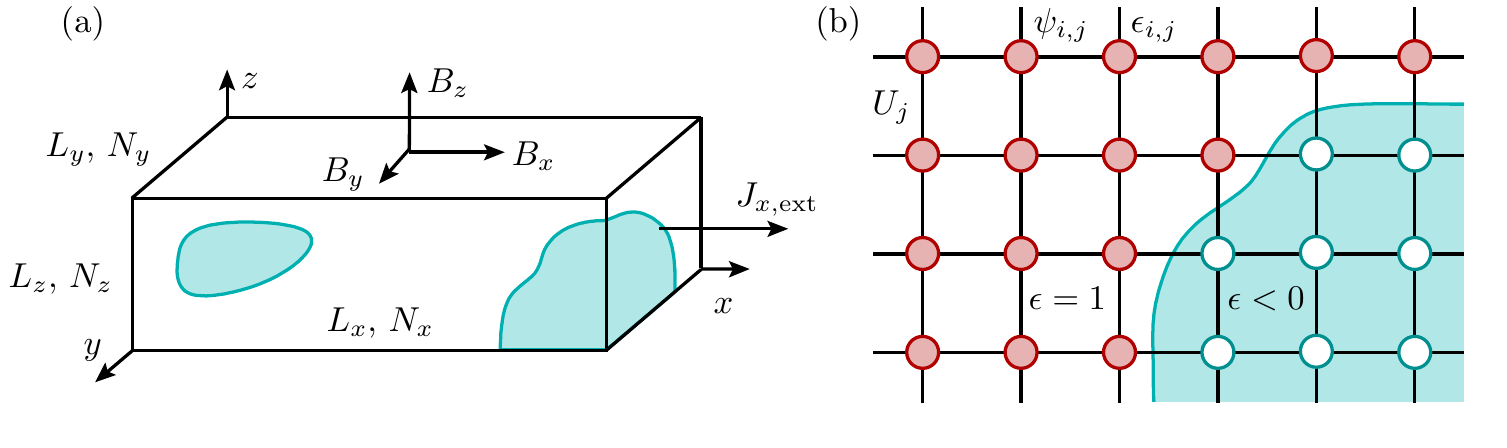}\label{fig:setup_a}}
        \subfloat{\label{fig:setup_b}}
    \end{center} \vspace{-4mm}
    \caption{
        (a)~Physical setup of the simulated system. The simulation cuboid has the size
        $L_x \times L_y \times L_z$. The magnetic field is applied either in $xz$ plane
        [Eq.~(\ref{eq:B_xz})] or $yz$ plane [Eq.~(\ref{eq:B_yz})].
        The external current is applied in $x$ direction.
        (b)~Sketch of a 2D cross section of the regular simulation mesh with fixed $z$-index $k$.
        The order parameter $\psi_{i,j,k}$ and $\Tc$ map $\epsilon_{i,j,k}$ are defined on 
        the grid points. Link variables~(\ref{eq:link_variables}) are defined on the edges 
        of the grid. The cyan grid points represent a non-superconducting defect, where 
        the critical temperature is below the actual temperature, $\epsilon_{i,j,k} < 0$.
    }
    \label{fig:setup}
\end{figure*}

Using these gauges, we can now generate expressions for the generalized gradient, $(\nabla - \imath\tilde{\bf A})\tpsi$, and Laplacian, $(\nabla -\imath\tilde{\bf A})^2\tpsi$, appearing in Eqs.~(\ref{eq:J_external_current}) and (\ref{eq:GL_external_current}), respectively. The gradient of $\tilde{\psi}$ is now explicitly
\begin{align*}
    (\nabla -\imath\tilde{\bf A}^{xz})\tpsi
    & = \bigl[ (\partial_x + \imath K + \imath B_z \bar y)\tpsi, \;
    \partial_y\tpsi, \;
    \nonumber \\ & \qquad\qquad
    (\partial_z - \imath B_x \bar y)\tpsi / g^2 \bigr]^{\tt T} \nonumber \\
    & = \bigl[ e^{-\imath x(K + \bar y B_z)}\partial_x (e^{\imath x(K + \bar y B_z)}\tpsi), \;
    \partial_y\tpsi,
    \nonumber \\ & \qquad\qquad
    e^{\imath \bar yzB_x}\partial_z (e^{-\imath \bar y z B_x}\tpsi)/g^2 \bigr]^{\tt T},
    \\
    (\nabla -\imath\tilde{\bf A}^{yz})\tpsi
    & = \bigl[ e^{-\imath x K}\partial_x (e^{\imath x K}\tpsi), \;
    e^{\imath \bar x y B_z} \partial_y (e^{-\imath \bar x y B_z} \tpsi),
    \nonumber \\ & \qquad\qquad
    e^{-\imath \bar x z B_y}\partial_z (e^{\imath \bar x z B_y}\tpsi)/g^2 \bigr]^{\tt T}
\end{align*}
for the magnetic field in the $xz$- and the $yz$ plane, respectively. The analogous expressions for the Laplacian are
\begin{subequations}
\begin{align}
    (\nabla -\imath\tilde{\bf A}^{xz})^2 \tpsi
    & = e^{-\imath x(K+\bar yB_z)}\partial_x^2 (e^{\imath x(K+\bar yB_z)}\tpsi)
    + \partial_y^2\tpsi
    \nonumber \\& \qquad
    + e^{\imath \bar yzB_x}\partial_z^2 (e^{-\imath \bar yzB_x}\tpsi) / g^2,
    \label{eq:LaplExp_xz}
    \\
    (\nabla -\imath\tilde{\bf A}^{yz})^2\tpsi
    & = e^{-\imath xK}\partial_x^2 (e^{\imath xK}\tpsi)
    \! + \! e^{\imath \bar xyB_z} \partial_y^2 (e^{-\imath \bar xyB_z} \tpsi)
    \nonumber \\& \qquad
    \! + \! e^{-\imath \bar xyB_y}\partial_z^2 (e^{\imath \bar xyB_y}\tpsi) / g^2.
    \label{eq:LaplExp_yz}
\end{align}
\end{subequations}
In the following sections we describe numerical implementation of solutions of Eqs.~(\ref{eq:GL_external_current}), (\ref{eq:Poisson}), and (\ref{eq:K_ODE}).

\section{Discretization of the coupled partial differential equations} \label{sec:discretization}

Solving for the evolving state of the system means integrating the GL equations Eq.~(\ref{eq:GL_external_current}) forward in time, then solving the Poisson equation Eq.~(\ref{eq:Poisson}) for the new scalar potential, and then integrating the ODE of Eq.~(\ref{eq:K_ODE}) forward in time for the electric field.

To do this, we introduce a discretization of the TDGL and Poisson equations that generates a numerically stable calculation that can easily be computationally implemented. We propose to use the symmetric implicit Crank-Nicolson integration scheme together with a linearization of the non-linear $|\psi|^2 \psi$-term of Eq.~(\ref{eq:GL_external_current}). We use this method instead of an explicit method, which would be vulnerable to numerical instabilities and inefficiency. And since we wish to accommodate both open and periodic boundary conditions, we also exclude using a quasi-spectral split step method based on fast Fourier transform (FFT), since it requires a completely or partially periodic domain on a structured grid. Furthermore, the use of internal boundaries to simulate insulating inclusions cannot be implemented using FFT.

For simplicity, we choose to model a domain with a cuboid shape of physical dimension $L_x \times L_y \times L_z$ in units of $\xi_0$ that is aligned with the Cartesian axes, as shown in Fig.~\subref{fig:setup_a}. This domain is discretized using a regular mesh with $N_x \times N_y \times N_z$ grid points, total $N=N_x N_y N_z$.

The coordinates are discretized as $x = i h_x,$ $y = j h_y$, and $z = k h_z$, where $i \in [0, N_x-1],$ $j \in [0, N_y-1] $, and $k \in [0, N_z-1]$. The dimensions of a  mesh element are  $h_x = L_x / (N_x-1)$, $h_y = L_y / (N_y-1)$, and $h_z = L_z / (N_y-1)$ in case of open boundary conditions (which will be discussed in Sec.~\ref{sec:bc_open} in details) or by $h_x = L_x / N_x$, $h_y = L_y / N_y$, and $h_z = L_z / N_z$ for quasi-periodic or periodic boundary conditions (Sec.~\ref{sec:bc_periodic}). We introduce the multi-index $m = (i,j,k) = i+N_x(j+kN_y) \in [0,N-1] $ as short notation for $i$, $j$, and $k$.

The value of the coefficient  $\epsilon({\bf r})$ is set at each spatial coordinate to model inclusions or different materials. For example, in Fig.~\subref{fig:setup_b}, the mesh points in the interior of a non-superconducting inclusion (open circles) have $\Tc({\bf r}) < T$, i.e., $\epsilon({\bf r}) < 0$, while the mesh points in the exterior superconducting material have $\epsilon({\bf r})=1$.

\subsection{Discretization of the Ginzburg-Landau equations} \label{sec:discretization_GL}

To solve the GL Eq.~(\ref{eq:GL_external_current}), we use the implicit Crank-Nicolson scheme for the time integration over a time step\footnote{The factor $u$ in Eq.~(\ref{eq:GL_external_current}) is absorbed in $h_t$.}~$h_t$
\begin{align}
    \tpsi_m-\tpsi_m^-
    & = h_t\Bigl[ (\epsilon_m \! - \! \imath u\tilde{\mu})\tpsi_m^{\Join}
    - \frac{|\tpsi_m|^2\tpsi_m}{2}
    - \frac{|\tpsi_m^-|^2\tpsi_m^-}{2}
    \nonumber \\&
    + ( \nabla \! - \! \imath\tilde{\bf A})^2\tpsi_m^{\Join}+\zeta_m(t)\Bigr],
    \label{eq:GL_numCN}
\end{align}
where $\tpsi_m=\tpsi_m(t)$, $\tpsi_m^-=\tpsi_m(t-h_t)$, and $\tpsi_m^{\Join} \equiv (\tpsi_m+\tpsi_m^-) /2$. This leads to a system of coupled non-linear equations. Therefore, we assume that $|\tpsi_m|^2$ does not change too much from one time step to the next and replace it by an estimate $|\tpsi_m^{\rm est}|^2$. Here we use $\tpsi_m^{\rm est} = \tpsi_m^-$.

To discretize the gauge invariant Laplacian in Eq.~(\ref{eq:GL_numCN}), we use the presentation (\ref{eq:LaplExp_xz})--(\ref{eq:LaplExp_yz}) and introduce the complex-valued link variables $\LV{\alpha}{\ell}$ defined on the edges of the simulations mesh, 
\begin{equation}
	\LV{\alpha}{\ell} = e^{-\imath h_{\alpha}{\tilde A}_{\alpha,\ell}},
	\label{eq:link_variables}
\end{equation}
where index $\ell = \{j, i\}$ defines type of link variable, $j$ is used for vector potential ${\bf\tilde A}^{xz}$  and $i$ is used for vector potential ${\bf\tilde A}^{yz}$. The index $\alpha = \{ x, y, z \}$ defines a component of ${\bf\tilde A}$. ${\tilde A}_{\alpha,\ell}$ are defined as
\begin{subequations}
\begin{align}
	{\tilde A}_{\alpha,j} & = {\tilde A}^{xz}_{\alpha}|_{y = jh_y}, \\
	{\tilde A}_{\alpha,i} & = {\tilde A}^{yz}_{\alpha}|_{x = ih_x}.
\end{align}
\end{subequations}
Using these link variables, the gauge-invariant Laplacian $(\nabla -\imath\tilde{\bf A})^2\tpsi_m$  can be discretized as central differences over the neighboring grid points,
\begin{subequations}
\begin{align}
    (\nabla - & \imath\tilde{\bf A}^{xz})^2\tpsi_m
    = (\LVU\tpsi_{i+1,j,k} + \LVU^*\tpsi_{i-1,j,k} \! - \! 2\tpsi_m) / h_x^2
        \nonumber \\ &
    + (\tpsi_{i,j+1,k} + \tpsi_{i,j-1,k}-2\tpsi_m) / h_y^2
        \nonumber \\ &
    + (\LVV\tpsi_{i,j,k+1} + \LVV^*\tpsi_{i,j,k-1}-2\tpsi_m) / g^2 h_z^2
    \label{eq:LaplUV_xz} \\
    (\nabla - & \imath\tilde{\bf A}^{yz})^2\tpsi_m
    = (\LVQ\tpsi_{i+1,j,k} + \LVQ^*\tpsi_{i-1,j,k} \! - \! 2\tpsi_m) / h_x^2
        \nonumber \\ &
    + (\LVW\tpsi_{i,j+1,k} + \LVW^*\tpsi_{i,j-1,k}-2\tpsi_m) / h_y^2
        \nonumber \\ &
    + (\LVZ\tpsi_{i,j,k+1} + \LVZ^*\tpsi_{i,j,k-1}-2\tpsi_m) / g^2 h_z^2.
    \label{eq:LaplUV_yz}
\end{align}
\end{subequations}

The linearized equation system for $\tPsi = \{ \tpsi_m \} = [\tpsi_0, \ldots, \tpsi_{N-1}]^{\tt T}$ can now be expressed as
\begin{equation}
    {\cal M} \tPsi = {\bf b},
    \label{eq:GL_linearized}
\end{equation}
where ${\cal M}$ is a sparse $N\times N$ matrix with non-zero elements given by (for the magnetic field in $xz$ plane)
\begin{align*}
    & {\cal M}_{(i,j,k),(i,j,k)} = 1 + (h_t/2) \bigl[
        |\tpsi_m^{\rm est}|^2 + \\ & \qquad\qquad\qquad
        2 (
            1/h_x^2 + 1/h_y^2 + 1/g^2 h_z^2
        ) + \imath u \tmu_m - \epsilon_m
    \bigr], \\
    & {\cal M}_{(i,j,k),(i+1,j,k) }
        = {\cal M}_{(i,j,k),(i-1,j,k)}^*
        = - h_t \LVU / 2 h_x^2, \\
    & {\cal M}_{(i,j,k),(i,j+1,k)}
        = {\cal M}_{(i,j,k),(i,j-1,k)}
        = - h_t / 2 h_y^2, \\
    & {\cal M}_{(i,j,k),(i,j,k+1)}
        = {\cal M}_{(i,j,k),(i,j,k-1)}^*
        = - \LVV h_t / 2 g^2 h_z^2,
\end{align*}
otherwise ${\cal M}_{m,m'} = 0$.

This matrix, in effect, applies a seven-point stencil around each grid point. The vector ${\bf b} = \{ b_m \} = [b_0, \ldots, b_{N-1}]^{\tt T}$ of the length $N$ is a function of the field at the prior time step and is defined as
\begin{align}
    b_m
    & = \tpsi_m^- + (h_t/2) \bigl[ (\epsilon_m \! - \! \imath u \tmu_m)\tpsi_m^- - \! |\tpsi_m^-|^2\tpsi_m^-
    \nonumber\\ & \quad
    + (\nabla \! - \! \imath\tilde{\bf A} )^2\tpsi_m^- + 2\zeta_m(t) \bigr].
    \label{eq:b}
\end{align}

Instead of inverting the linear system of equation in Eq.~(\ref{eq:GL_linearized}), the solution can be found via an iterative Jacobi method. The matrix ${\cal M}$ can be split into two parts, ${\cal M} = {\cal D} + {\cal R}$, where ${\cal D} = {\rm diag} {\cal M}$. For the Jacobi method one typically uses ${\cal D}$ as left preconditioner, and solves the equation
\begin{equation*}
    {\cal\hat M} \tPsi = {\bf\hat b},
\end{equation*}
where ${\cal\hat M} = {\cal D}^{-1} {\cal M}$ and ${\bf\hat b} = {\cal D}^{-1}{\bf b}$. With ${\cal\hat R}={\cal D}^{-1} {\cal R}$ this results in the equation
\begin{equation}
    \tPsi = {\bf\hat b}-{\cal\hat R}\tPsi.
\end{equation}
The iterative Jacobi scheme is then given by
\begin{equation*}
    \tPsi^{(l+1)} = {\bf\hat b} - {\cal\hat R} \tPsi^{(l)},
\end{equation*}
where $\tPsi^{(l)}$ are successively better approximations for the solution of Eq.~(\ref{eq:GL_linearized}). As the matrix ${\cal\hat M}$ is diagonally dominant the convergence of this iteration is guaranteed. Writing this scheme in terms of components of the original ${\cal M}$ and ${\bf b}$, we get
\begin{equation}
    \tpsi_m^{(l+1)}
    = \frac{1}{{\cal M}_{m,m}}
    \Bigl(b_m - \sum\limits_{\substack{m' = 0 \\ m' \neq m}}^N {\cal M}_{m,m'} \, \tpsi_{m'}^{(l)}\Bigr).
    \label{eq:final}
\end{equation}
Defining the residual ${\bf r}^{(l)} = {\bf b} - {\cal M} \tPsi^{(l)}$ as the deviation of $\tPsi^{(l)}$ from the real solution, the convergence criterion is given by 
\begin{equation}
	\Vert {\bf r}^{(l)} \Vert_{\max}^2
	< \varepsilon.
	\label{eq:convergence}
\end{equation}

\subsubsection{Current-voltage characteristics}

To solve the ODE for the electric field given in Eq.~(\ref{eq:K_ODE}), we calculate the averaged quantity $J_{\rm av}$ defined in Eq.~(\ref{eq:J_av}). The gauge-invariant gradient $(\nabla -\imath\tilde{\bf A})\tpsi_m$ is approximated in Sec.~\ref{sec:large_lambda} and discretized as central differences over the neighboring grid points using the link variables,
\begin{equation}
    J_{\rm av}
    = \frac{1}{2N h_x} \sum\limits_{m=0}^{N-1}
        {\rm Im} [
            \tpsi_m^* (\LVU \tpsi_{i+1,j,k} \! - \! \LVU^* \tpsi_{i-1,j,k} )
        ]
\end{equation}
for a magnetic field in the $xz$ plane; for a magnetic field in $yz$ plane, simply replace $\LVU$  by $\LVQ$.

\subsubsection{Noise terms}

The definition of the noise terms in Eq.~(\ref{eq:b}) and the current fluctuation for a spatial and temporal discretization of a domain is subtle. Since we have to relate the numerical noise amplitude to real temperature, we go back to dimensional units and distinguish these from dimensionless units explicitly; the latter are denoted by overbars.

The simplest way to obtain noise in the reduced units is to consider an equation, valid at short times, describing the diffusion of the order parameter
\begin{equation}
    \Gamma \partial_t \psi ({\bf r},t) = \zeta ({\bf r},t).
    \label{eq:noise_dynamics}
\end{equation}
A dimensionless version of the noise term $\bar \zeta (\bar {\bf r},\bar t)$ can be obtained by rewriting this equation as
$
    ({\Gamma\psi_0}/{t_0}) \, \partial_{\bar t} \bpsi ({\bf r},t)
    = \zeta ({\bf r},t),
$
where $\bar{\psi}$ is the dimensionless order parameter and $t_0 = 4\pi\sigma\lambda_0^2/c^2$. Introducing
$
    \bar\zeta
    = {t_0 u \zeta}/{\Gamma\psi_0}
    = {\zeta}/{a_0 \psi_0}
$
results in
\begin{equation}
    u \partial_{\bar t} \bpsi ({\bf r},t)
    = \bar\zeta ({\bf r},t),
    \label{eq:noise_dynamics_r}
\end{equation}
where $u = \Gamma/a_0 t_0$. The noise correlator in Eq.~(\ref{eq:noise_zeta}) can now be rewritten as
\begin{equation}
    \langle\bar\zeta^*({\bf \bar r},\bar t) \bar\zeta({\bf \bar r}',\bar t')\rangle
    = u \bar T \delta({\bf \bar r} - {\bf \bar r}' ) \delta(\bar t - \bar t'),
\end{equation}
where $\bar T= {2 T}/{a_0\psi_0^2\xi_0^d}$ and $d = 2$, $3$ is the dimensionality of the system. Since the condensation energy per unit volume is
$
    E_{k0}
    = {a_0 \psi_0^2}/{2},
$
the reduced temperature is $\bar T = T/(E_{k0}\xi_0^d)$. Thus, in dimensionless units $\bar t = t/t_0$ and $\bar {\bf r} = {\bf r}/\xi_0$, we get $\bpsi(\bar t) = ({1}/{u}) \int\nolimits_0^{\bar t}d\bar t_1\,\bar \zeta(\bar t_1)$ with $\bpsi(0)=0$. The dynamics of $\psi$ is given by the correlator
\begin{align*}
    G(\bar {\bf r}, \bar {\bf r}', \bar t)
    & = \langle \bpsi^*(\bar {\bf r},\bar t) \bpsi(\bar {\bf r}',\bar t) \rangle \\
    & \!\!\!\! = \frac{1}{u^2}
    	\int\nolimits_0^{\bar t} \! {\rm d}{\bar t}_1 {\rm d}{\bar t}_2 \,
    	\langle\bar\zeta^*(\bar {\bf r},\bar t_1) \bar\zeta(\bar {\bf r}',\bar t_2) \rangle
    = \frac{\bar T \bar t}{u} \delta(\bar {\bf r} - \bar {\bf r}' ).
\end{align*}

Numerically, we model thermal noise as independent complex terms $\bar\zeta = \bar\zeta_1 + \imath\bar\zeta_2$ for every mesh point and time step, in which $\bar\zeta_1 $ and $\bar\zeta_2$ are random numbers uniformly distributed over the range $[-{\bar\zeta}_{\rm max},\,{\bar\zeta}_{\rm max}]$. To establish a relation between the numerical parameter ${\bar\zeta}_{\rm max}$ and the physical parameter~$\bar T$, we calculate the order parameter correlator for the discrete time version of Eq.~(\ref{eq:noise_dynamics_r}). Using the formal solution of this equation
\begin{equation*}
    \bpsi(\bar t)
    = \frac{h_{\bar t}}{u} \sum\limits_{n=0}^{N_t} \bar\zeta(n h_{\bar t}),
\end{equation*}
where ${\bar t} = N_t h_{\bar t}$, and the noise correlator
\begin{align*}
    \langle\bar\zeta^*(\bar{\bf r}_i, n_1 h_{\bar t})
        \bar\zeta(\bar{\bf r}_j, n_2 h_{\bar t}) \rangle
    & = \delta_{n_1,n_2}\delta_{i,j} \langle\bar\zeta_1^2 + \bar\zeta_2^2\rangle \\
    & = (2/3) \delta_{n_1,n_2}\delta_{i,j} {\bar\zeta}_{\rm max}^2,
\end{align*}
we obtain
\begin{equation*}
    {\bar G}_{i,j}
    = \frac{2}{3}{\bar\zeta}_{\rm max}^2\delta_{i,j}\frac{h_{\bar t}^2}{u^2}
        \sum\limits_{n_1=0}^{N_t} \sum\limits_{n_2=0}^{N_t} \delta_{n_1,n_2}
    = \frac{2}{3}{\bar\zeta}_{\rm max}^2\delta_{i,j}\frac{h_{\bar t} \bar t}{u^2}
\end{equation*}
for the discrete correlation function of the order parameter. Comparing both expressions for $G$ and using the discretization $\delta(\bar {\bf r} - \bar {\bf r}') = \delta_{i,j} / h_x h_y h_z$, we obtain
$
    {2{\bar\zeta}_{\rm max}^2 h_{\bar t}} / {3 u^2}
    = {\bar T} / {u h_x h_y h_z}
$
or
\begin{align}
    {\bar\zeta}_{\rm max} 
    & = ( {3\bar T u}/{2 h_{\bar t} h_x h_y h_z} )^{1/2} \nonumber \\
    & = ( {3 T b u}/{h_{\bar t} h_x h_y h_z a_0^2\xi_0^d} )^{1/2}.
\end{align}
This expression implies that ${\bar\zeta}_{\rm max}$ depends on the grid precision and time discretization as ${\bar\zeta}_{\rm max}^2 \propto (h_{\bar t} h_x h_y h_z)^{-1}$. Therefore, we use the parameter ${\bar T}_\zeta = h_t h_x h_y h_z {\bar\zeta}_{\rm max}^2$ as the grid-independent simulation parameter that characterizes the thermal noise in the system. The current noise term ${\cal I}({\bf r},t)$ can be analyzed in a similar way.

\subsection{Discretization of the Poisson equation} \label{sec:discretization_Poisson}

The differential Eq.~(\ref{eq:Poisson}) for $\tmu$ is discretized as follows: The left side is discretized as
\begin{align*}
    & (\tmu_{i+1,j,k}+\tmu_{i-1,j,k}-2\tmu_m) / h_x^2 +
    \nonumber \\
    & (\tmu_{i,j+1,k}+\tmu_{i,j-1,k}-2\tmu_m) / h_y^2 +
    \nonumber \\
    & (\tmu_{i,j,k+1}+\tmu_{i,j,k-1}-2\tmu_m) / g^2 h_z^2,
\end{align*}
while the right side is discretized as
\begin{align*}
    & (1 / h_x^2) \, {\rm Im} [
        \tpsi_m^*
        (\LVU\tpsi_{i+1,j,k} + \LVU^*\tpsi_{i-1,j,k} )
    ] +
    \nonumber \\ &
    (1 / h_y^2) \, {\rm Im} [
        \tpsi_m^*
        (\tpsi_{i,j+1,k} \! + \! \tpsi_{i,j-1,k} )
    ] +
    \nonumber \\ &
    (1 / g^2 h_z^2) \, {\rm Im} [
        \tpsi_m^*
        (\LVV\tpsi_{i,j,k+1} \! + \! \LVV^*\tpsi_{i,j,k-1} )
    ]
\end{align*}
for a magnetic field in $xz$ plane. For the field the $yz$ plane, the link variables should be replaced as $\LV{\alpha}{j} \to \LV{\alpha}{i}$. To solve this elliptical PDE, we use an over-relaxation method. We introduce a virtual relaxation time~$\tau$ and iteratively solve the following differential equation
\begin{equation*}
    \partial_\tau \tmu
    = - \Delta \tmu
    + {\bf \nabla} \, {\rm Im} [ \tpsi^*({\bf\nabla}-\imath\tilde{\bf A})\tpsi ].
\end{equation*}
Again we use the implicit Crank-Nicolson scheme
\begin{align*}
    \tmu_m \! - \!  \tmu_m^-
    & = h_\tau \bigl\{
    - (\tmu_{i+1,j,k}^{\Join}+\tmu_{i-1,j,k}^{\Join}-2\tmu_m^{\Join}) / h_x^2
    \nonumber \\ &
    - (\tmu_{i,j+1,k}^{\Join}+\tmu_{i,j-1,k}^{\Join}-2\tmu_m^{\Join}) / h_y^2
    \nonumber \\ &
    - (\tmu_{i,j,k+1}^{\Join}+\tmu_{i,j,k-1}^{\Join}-2\tmu_m^{\Join}) / g^2 h_z^2
    \nonumber \\ &
    + (1 / h_x^2) \, {\rm Im} [
        \tpsi_m^*
        (\LVU\tpsi_{i+1,j,k} + \LVU^*\tpsi_{i-1,j,k} )
    ]
    \nonumber \\ &
    + (1 / h_y^2) \, {\rm Im} [
        \tpsi_m^*
        (\tpsi_{i,j+1,k}
        + \tpsi_{i,j-1,k} )
    ]
    \nonumber \\ &
    + (1 / g^2 h_z^2) \, {\rm Im} [
        \tpsi_m^*
        (\LVV\tpsi_{i,j,k+1} \! + \! \LVV^*\tpsi_{i,j,k-1} )
    ]
    \bigr\},
\end{align*}
where $h_\tau$ is the virtual time step, $\tmu_m^- = \tmu_m(\tau-h_\tau)$ is the value at the previous step, and $\tmu_m^{\Join} = (\tmu_m + \tmu_m^-)/2$. So, at each time step we solve the linear matrix equation
\begin{equation}
    {\cal N}\tmu = {\bf c},
    \label{eq:poissonlinear}
\end{equation}
where ${\cal N}$ is a sparse $N\times N$ matrix with elements
\begin{align*}
    & {\cal N}_{(i,j,k),(i,j,k)} = 1 + h_\tau ( 1/h_x^2 + 1/h_y^2 + 1/g^2 h_z^2 ), \\
    & {\cal N}_{(i,j,k),(i+1,j,k) } = {\cal N}_{(i,j,k),(i-1,j,k) }
    = - h_\tau / 2h_x^2, \\
    & {\cal N}_{(i,j,k),(i,j+1,k) } = {\cal N}_{(i,j,k),(i,j-1,k) }
    = - h_\tau / 2h_y^2, \\
    & {\cal N}_{(i,j,k),(i,j,k+1) } = {\cal N}_{(i,j,k),(i,j,k-1) }
    = - h_\tau / 2 g^2 h_z^2,
\end{align*}
(otherwise ${\cal N}_{m,m'} = 0$), that applies a seven-point stencil around each grid point. Vector ${\bf c} = \{ c_m \} = [c_0, \ldots, c_{N-1}]^{\tt T}$ of the length $N$ is a function of $ \tmu$ at the prior time step and is defined as
\begin{align*}
    c_m & = \tmu_m^-
    + h_\tau \bigl\{
    - ( \tmu_{i+1,j,k}^- \! + \! \tmu_{i-1,j,k}^- \! - \! 2\tmu_m^- ) / 2 h_x^2
    \nonumber \\ &
    - ( \tmu_{i,j+1,k}^- \! + \! \tmu_{i,j-1,k}^- \! - \! 2\tmu_m^- ) / 2 h_y^2
    \nonumber \\ &
    - ( \tmu_{i,j,k+1}^-+\tmu_{i,j,k-1}^--2\tmu_m^- ) / 2 g^2 h_z^2
    \nonumber \\ &
    + (1 / h_x^2) \, {\rm Im} [
        \tpsi_m^*
        (\LVU\tpsi_{i+1,j,k} + \LVU^*\tpsi_{i-1,j,k} )
    ] \nonumber \\
    & + (1 / h_y^2) \, {\rm Im} [
        \tpsi_m^*
        (\tpsi_{i,j+1,k} \! + \! \tpsi_{i,j-1,k} )
    ]
    \nonumber \\ &
    + (1 / g^2 h_z^2) \, {\rm Im} [
        \tpsi_m^*
        (\LVV\tpsi_{i,j,k+1} \! + \! \LVV^*\tpsi_{i,j,k-1} )
    ]
    \bigr\}.
\end{align*}

Without the relaxation, the diagonal elements of the discretized linear equation system are of the same order as the off-diagonal elements leading to poor convergence of the iterative solver.

A formulation like that in Eq.~(\ref{eq:final}) is implemented to iteratively solve the linear systems of equations. We solve this equations each real time step until the convergence criterion $||\partial_\tau \tmu||_{\max}^2 < \varepsilon_\mu$ is satisfied.

\subsection{Boundary conditions} \label{sec:boundary_conditions}

Two different types of boundary condition can be implemented at each of the three boundaries of the cuboidal domain (that is, in the $x$, $y$, and $z$ direction). The simplest boundary condition to implement is a periodic boundary condition where quasi-periodic conditions are not necessary. Here, identical calculations are performed at the end of the grid as in the middle, (e.g. with the index convention: $i=-1 \Rightarrow i = N_x-1$ and $i=N_x \Rightarrow i = 0$) and nothing more needs to be said. However, if quasi-periodic conditions do apply, then a calculation of the phase jump at the boundary is necessary. This is discussed in Sec.~\ref{sec:bc_periodic}. The second type of boundary condition is an open (Neumann) boundary condition which assumes that  the current normal to the boundary vanishes. Implementation of this type of boundary condition is discussed in Sec.~\ref{sec:bc_open}.

\subsubsection{Quasi-periodic boundary conditions} \label{sec:bc_periodic}

For the two Landau-gauge vector potentials presented in Sec.~\ref{sec:large_lambda}, the linear $y$ and $x$ dependence causes the order parameter $\psi$ to be quasi-periodic in the $y$ and $x$ direction, respectively. Quasi-periodic means that only the amplitude of the order parameter is periodic, and therefore continuous at the boundary, while its phase can have a discontinuity, that is, a phase jump. Note that the value of this phase jump varies over the boundary
surface.

Quasi-periodic boundary conditions for the order parameter $\tpsi$ can be derived by equating the amplitudes $|\tpsi|_{x=0} = |\tpsi|_{x=L_x}$, $|\tpsi|_{y=0} = |\tpsi|_{y=L_y}$, $|\tpsi|_{z=0} = |\tpsi|_{z=L_z}$ and ensuring the components of the current $\bf J$ are continuous across the boundaries ${\bf J}|_{x=0} = {\bf J}|_{x=L_x}$, ${\bf J}|_{y=0} = {\bf J}|_{y=L_y}$, ${\bf J}|_{z=0} = {\bf J}|_{z=L_z}$. The scalar potential $\tmu$ is always continuous at the boundary, i.e., $\tmu_{x=0} = \tmu_{x=L_x}$, $\tmu_{y=0} = \tmu_{y=L_y}$, $\tmu_{z=0} = \tmu_{z=L_z}$, since it has no phase and by its construction in Eq.~(\ref{eq:mu_transformation}).

\paragraph{Magnetic field in the $xz$ plane, quasi-periodic in the $y$ direction.} To solve for the phase jump, defined through a complex phase factor $P_{i,k}$, of $\tpsi$ across the $y$ boundary, such that $\tpsi _{i,-1,k}= P_{i,k}\tpsi_{i,N_y-1,k}$ we use $|\tpsi_{i,-1,k}| =| \tpsi_{i,N_y-1,k}|$ (continuity of the magnitude of $\psi$) and  $\{(\nabla -\imath\tilde{\bf A})\tpsi_{i,-1,k}\}_x = \{(\nabla - \imath\tilde{\bf A})\tpsi_{i,N_y-1,k}\}_x$ and $\{(\nabla -\imath\tilde{\bf A})\tpsi_{i,-1,k}\}_y = \{(\nabla - \imath\tilde{\bf A})\tpsi_{i,N_y-1,k}\}_y$  (continuity of ${\bf J}$). The first gradient equivalence can be expressed as
\begin{align*}
    & \tpsi_{i,N_y-1,k}^* [\LV{x}{j=N_y-1} \tpsi_{i+1,N_y-1,k} - \LV{x}{j=N_y-1}^* \tpsi_{i-1,N_y-1,k}]
    \\& \quad
    = \tpsi_{i,-1,k}^* [\LV{x}{j=-1} \tpsi_{i+1,-1,k} - \LV{x}{j=-1}^* \tpsi_{i-1,-1,k}] \\
    & \quad = P_{i,k}^* \tpsi_{i,N_y-1,k}^* [\LV{x}{j=-1} P_{i+1,k} \tpsi_{i+1,N_y-1,k}
    \\& \qquad
    - \LV{x}{j=-1}^* P_{i-1,k} \tpsi_{i-1,N_y-1,k}].
\end{align*}
Using $\LV{x}{j=-1} = e^{\imath h_x(K-(h_y + L_y/2) B_z)}$ and $\LV{x}{j=N_y-1} = e^{\imath h_x(K-(h_y - L_y/2) B_z)} $ [see Eq.~(\ref{eq:link_variables})], we have $P_{i,k}^* P_{i+1,k} e^{-\imath h_x L_y B_z} = 1$ and $P_{i,k}^* P_{i-1,k} e^{\imath h_x L_y B_z} = 1$. As a result,
$
    P_{i,k} = e^{\imath i h_x L_y B_z}p_k,
$
where $p_k$ is the $k$-dependent part.

The second gradient equivalence can be expressed as
\begin{align*}
    & \tpsi_{i,N_y-1,k}^* [\LV{z}{j=N_y-1} \tpsi_{i,N_y-1,k+1} - \LV{z}{j=N_y-1}^* \tpsi_{i,N_y-1,k-1}] \\
    & \quad = \tpsi_{i,-1,k}^* [\LV{z}{j=-1} \tpsi_{i,-1,k+1} - \LV{z}{j=-1}^* \tpsi_{i,-1,k-1}] \\
    & \quad = P_{i,k}^* \tpsi_{i,N_y-1,k}^* [\LV{z}{j=-1} P_{i,k+1} \tpsi_{i,N_y-1,k+1}
    \\& \qquad
    - \LV{z}{j=-1}^* P_{i,k-1} \tpsi_{i,N_y-1,k-1}].
\end{align*}
Using $\LV{z}{j=-1} = e^{\imath h_z(h_y + L_y/2) B_x}$ and $\LV{z}{j=N_y-1} = e^{\imath h_z(h_y - L_y/2) B_x}$ [see Eq.~(\ref{eq:link_variables})], we have $P_{i,k}^* P_{i,k+1} e^{\imath h_z L_y B_x} = 1$ and $P_{i,k}^* P_{i,k-1} e^{-\imath h_z L_y B_x} = 1$. Thus, $p_k = e^{-\imath k h_z L_y B_x}$ and
\begin{equation*}
    P_{i,k} = e^{\imath (i h_x L_y B_z - k h_z L_y B_x)}.
    \label{eq:qp_phasejump}
\end{equation*}
The second derivative in the $y$ directions at $j = 0$ and $j = N_y-1$ [Eq.~(\ref{eq:LaplUV_xz}) can be rewritten in terms of indices inside the simulation domain,
\begin{align*}
    (\partial_y - \imath\tilde A_y)^2 \tpsi|_{y=0}
    & = (\tpsi_{i,1,k} + P_{i,k} \tpsi_{i,N_y-1,k} \nonumber\\& \qquad
        - 2\tpsi_{i,0,k})/h_y^2, \\
    (\partial_y - \imath\tilde A_y)^2 \tpsi|_{y=L_y}
    & = (P_{i,k}^*\tpsi_{i,0,k} + \tpsi_{i,N_y-2,k} \nonumber\\& \qquad
        - 2\tpsi_{i,N_y-1,k})/h_y^2.
\end{align*}

\paragraph{Magnetic field in $yz$ plane, quasi-periodic in the $x$ direction.} To write the quasi-periodic boundary conditions at $x = 0$ and $x = L_x$ we perform the same operations as in above. We have $\tpsi_{-1,j,k} = P_{j,k} \tpsi_{N_x-1,j,k}$, where
\begin{equation*}
    P_{j,k} = e^{\imath (-j h_y L_x B_z + k h_z L_x B_y)}.
\end{equation*}
The second derivative in $x$ direction at $i = 0$ and $i = N_x-1$ [Eq.~(\ref{eq:LaplUV_yz})] in terms of indices inside the simulation domain is
\begin{align*}
    (\partial_x - \imath\tilde A_x)^2\tpsi |_{x=0}
    & = (\LVQ \tpsi_{1,j,k} + \LVQ^* P_{j,k}\tpsi_{N_y-1,i,k} \nonumber\\& \qquad
        - 2\tpsi_{0,i,k})/h_x^2, \\
    (\partial_x - \imath\tilde A_x)^2\tpsi |_{x=L_x}
    & = (\LVQ P_{j,k}^*\tpsi_{0,j,k} + \LVQ^* \tpsi_{N_x-2,j,k} \nonumber\\& \qquad
        - 2\tpsi_{N_y-1,j,k})/h_x^2.
\end{align*}

\paragraph{Implications for the magnetic field.} If the system is periodic in two or all three directions, then the quasi-periodic boundary condition places restrictions on the choice of $B_x$, $B_y$, and $B_z$ that can be used. However, this restriction is not severe. Namely, the phase jump $P_{i,k}$ (or $P_{j,k}$) must itself be continuous in the $x,z$ (or $y,z$) directions. Thus, for an $xz$ plane magnetic field, both the $x$ and $z$-component of the magnetic field must integer multiples of the magnetic flux in the corresponding cross-sections, or  $B_x = 2\pi n_x/(L_y L_z)$ and $B_z = 2\pi n_z/(L_x L_y)$, where  $n_x = 0,\,\pm 1,\,\pm 2,\,\ldots$ and  $n_z = 0,\,\pm 1,\,\pm 2,\,\ldots$ For moderately sized cuboid domains, the possible choice of $B_x$ and $B_z$ is still near continuous.

\subsubsection{Open boundary conditions} \label{sec:bc_open}

Open (Neumann) boundary conditions imply a zero current perpendicular to the boundary surface ${\bf n J} = 0$ or, equivalently, ${\bf n} (\nabla - \imath\tilde{\bf A}) \tpsi = 0$, where $\bf n$ is the unit normal vector. Below, we explicitly write the discretized versions of the gradients and Laplacians at the boundary.

\paragraph{$x$ direction.} The boundary conditions at the $x = 0$ and $x = L_x$ surfaces have the form $(\partial_x - \imath\tilde A_x)\tpsi|_{x=0,L_x}=0$. For a magnetic field in the $xz$ plane in discrete form, using link variables, this requires $\LVU^*\tpsi_{-1,j,k} = \LVU\tpsi_{1,j,k}$ and $\LVU\tpsi_{N_x,j,k} = \LVU^*\tpsi_{N_x-2,j,k}$ at $i = 0$ and correspondingly at $i = N_x$. Similarly, for a magnetic field in the $yz$ plane this requires $\LVQ^*\tpsi_{-1,j,k} = \LVQ\tpsi_{1,j,k}$ and $\LVQ\tpsi_{N_x,j,k} = \LVQ^*\tpsi_{N_x-2,j,k}$ at $i = 0$ and $i = N_x$ correspondingly. Accordingly, the second derivative at these surfaces are given by
\begin{align*}
    (\partial_x - \imath\tilde A_x^{xz})^2\tpsi|_{x=0}
    & = 2(\LVU\tpsi_{1,j,k}-\tpsi_{0,j,k})/h_x^2, \\
    (\partial_x - \imath\tilde A_x^{xz})^2\tpsi|_{x=L_x}
    & = 2(\LVU^*\tpsi_{N_x-2,j,k}-\tpsi_{N_x-1,j,k})/h_x^2
\end{align*}
and
\begin{align*}
    (\partial_x - \imath\tilde A_x^{yz})^2\tpsi|_{x=0}
    & = 2(\LVQ\tpsi_{1,j,k}-\tpsi_{0,j,k})/h_x^2, \\
    (\partial_x - \imath\tilde A_x^{yz})^2\tpsi|_{x=L_x}
    & = 2(\LVQ^*\tpsi_{N_x-2,j,k}-\tpsi_{N_x-1,j,k})/h_x^2
\end{align*}
for Eq.~(\ref{eq:LaplUV_xz}) and (\ref{eq:LaplUV_yz}), respectively. Obviously, open boundary conditions in the $x$ direction cannot be used in conjunction a non-zero external current~$J_{x,{\rm ext}}$.

\paragraph{$y$ direction.} The corresponding boundary conditions at $y = 0$ and $y = L_y$ are $(\partial_y - \imath\tilde A_y)\tpsi|_{y=0,L_y}=0$. For a magnetic field in the $xz$ plane $\tpsi_{i,-1,k} = \tpsi_{i,1,k}$ and $\tpsi_{i,N_y,k} = \tpsi_{i,N_y-2,k}$ and for a magnetic field in the $yz$ plane $\LVW^*\tpsi_{i,-1,k} = \LVW\tpsi_{i,1,k}$ and $\LVW\tpsi_{i,N_y,k} = \LVW^*\tpsi_{i,N_y-2,k}$. The second derivative is written as
\begin{align*}
    (\partial_y - \imath\tilde A_y^{xz})^2\tpsi|_{y=0}
    & = 2(\tpsi_{i,1,k}-\tpsi_{i,0,k})/h_y^2, \\
    (\partial_y - \imath\tilde A_y^{xz})^2\tpsi|_{y=L_y}
    & = 2(\tpsi_{i,N_y-2,k}-\tpsi_{i,N_y-1,k})/h_y^2
\end{align*}
and
\begin{align*}
    (\partial_y - \imath\tilde A_y^{yz})^2\tpsi|_{y=0}
    & = 2(\LVW\tpsi_{i,1,k}-\tpsi_{i,0,k})/h_y^2, \\
    (\partial_y - \imath\tilde A_y^{yz})^2\tpsi|_{y=L_y}
    & = 2(\LVW^*\tpsi_{i,N_y-2,k}-\tpsi_{i,N_y-1,k})/h_y^2
\end{align*}
for Eq.~(\ref{eq:LaplUV_xz}) and (\ref{eq:LaplUV_yz}), respectively.

\paragraph{$z$ direction.} At $z = 0$ and $z = L_z$ boundary conditions are $(\partial_z - \imath\tilde A_z)\tpsi|_{z=0,L_z}=0$. For a magnetic field in the $xz$ plane $\LVV^*\tpsi_{i,j,-1} = \LVV\tpsi_{i,j,1}$ and $\LVV\tpsi_{i,j,N_z} = \LVV^*\tpsi_{i,j,N_z-2}$ and for a magnetic field in the $yz$ plane $\LVZ^*\tpsi_{i,j,-1} = \LVZ\tpsi_{i,j,1}$ and $\LVZ\tpsi_{i,j,N_z} = \LVZ^*\tpsi_{i,j,N_z-2}$. The second derivative is given by
\begin{align*}
    (\partial_z - \imath\tilde A_z^{xz})^2\tpsi|_{z=0}
    & = 2(\LVV \tpsi_{i,j,1}-\tpsi_{i,j,0})/h_z^2, \\
    (\partial_z - \imath\tilde A_z^{xz})^2\tpsi|_{z=L_z}
    & = 2(\LVV^*\tpsi_{i,j,N_z-2}-\tpsi_{i,j,N_z-1})/h_z^2
\end{align*}
and
\begin{align*}
    (\partial_z - \imath\tilde A_z^{yz})^2\tpsi|_{z=0}
    & = 2(\LVZ \tpsi_{i,j,1}-\tpsi_{i,j,0})/h_z^2, \\
    (\partial_z - \imath\tilde A_z^{yz})^2\tpsi|_{z=L_z}
    & = 2(\LVZ^*\tpsi_{i,j,N_z-2}-\tpsi_{i,j,N_z-1})/h_z^2
\end{align*}
for Eq.~(\ref{eq:LaplUV_xz}) and (\ref{eq:LaplUV_yz}), respectively.

\subsection{Pinning and geometry} \label{sec:pattern}

Perhaps the most scientifically attractive problem that can be modeled  by the system of equations above is the interaction between vortices and vortex-pinning defects in a superconductor under different conditions. Especially of interest are hybrid superconducting/non-superconducting defect structures in which one or more lattices of non-superconducting defects are embedded in a superconducting matrix. In the discretization of the equations above, a pinning landscape can be implemented in the following ways: (i)~by spatial  modulation of $\epsilon({\bf r})$ (via the choice of critical temperature) at each grid point representing metallic inclusions, and (ii)~by applying no-current (open) boundary conditions across internal grid edges simulating insulating inclusions.

Using a pattern generator, we can create the most common types of hybrid structures that are found in experiments and industrial applications, combining an arbitrary number of different patterns for both two- and three-dimensional domains. The geometry of defects that can be modeled include, but are not limited to, rectangles, cuboids, circles, spheres, ellipsoids, crosses, octants, and cylinders. These defects can be arranged in square, triangular, or honeycomb lattices in 2D, rectangular or hexagonal lattices in 3D, or random lattices in any dimension.

We can also model tessellations of different structures, such as a checkerboard, a standard Voronoi tessellation, based on the standard Euclidean distance function, or extensions of the standard Voronoi tessellation where different types of distance functions are used, e.g. absolute or maximum. For example, we can simulate a polycrystalline thin superconducting film with variations of~$\Tc$ in each crystallite using this method.

Also, rather than explicitly define a more complex non-cuboidal domain, different shapes of integration domains can be realized by simply imposing low/zero critical temperatures values at grid points around a desired shaped domain. For example, to study the magnetic angular dependence in finite size samples, a pattern of $\epsilon({\bf r})$ can be imposed to create an axially symmetric, cylindrical simulation domain.

\section{Implementation as a flexible simulation tool} \label{sec:implementation}

\subsection{Implementation on a GPU} \label{sec:GPU}

The discretization and integration scheme described above is easily amen\-able to implementing in a massively parallel multi-threaded environment. We have implemented a prototype of this algorithm on an NVIDIA GPU using the Compute Unified Device Architecture (CUDA).

If the total number grid points in the system is $N = N_x N_y N_z$, then the computation requires between seven and eleven arrays of floats (or doubles) of length $N$. These arrays are: The (i)~real and (ii)~imaginary components of the order parameter ${\tilde \psi}$, plus a second copy (iii)--(iv), the (v)~real and (vi)~imaginary components of the complex vector ${\bf b}$, (vii)~the real inverse diagonal matrix ${\cal D}^{-1} = {\rm diag}\{{\cal M}^{-1}_{0,0}, \ldots, {\cal M}^{-1}_{N-1,N-1}\}$. If a current is applied, then the computation also requires (viii)~an array for $\tmu$ plus a copy (ix), and (x) an array for the imaginary part of ${\cal D}^{-1}$. If inclusions are to be modeled via a modulated critical temperature, then, finally, (xi)~a spatially dependent $\epsilon({\bf r})$ must be stored in memory. 

These arrays need to be stored in the global memory of the GPU card and determines the maximal size of the computational mesh. In most cases, single precision is sufficient for the solver, particularly in the case when thermal noise is present. Thus we can simulate a square 2D system with up to $12\,188^2$ mesh points or a cubic 3D system with up to $548^3$ mesh points using single precision NVIDIA GPU with 6\,GB of memory. The spatial resolution is typically chosen as two grid points per coherence length, which translates to real system sizes of $\sim (6\,000\xi_0)^2$ in 2D superconducting films or $\sim (270\xi_0)^3$ for 3D cubic superconductors.

To advance a single time step, first, the next value of $\tpsi_m$ is solved for, then the electric field is re-calculated, then the next value for scalar potential~$\tmu$ is solved. The algorithm for solving for $\tpsi_m$ is divided into two kernel functions. The first kernel initializes the array ${\cal D}^{-1}$ and the vector~${\bf b}$ of Eq.~(\ref{eq:final}), taking into account all boundary conditions, thermal noise, and the pinning landscape. The second kernel implements the iteration step and convergence check. The calculation of the non-zero off-diagonal elements ${\cal M}_{m,m'}$ is performed on the fly inside the kernel. One thread is assigned to each grid point of the mesh, or $\tpsi_m$ calculation. To avoid performing a summation over all the threads calculations for the convergence check, a maximum norm convergence check is used, where all the threads must agree that their calculations have converged before the iteration terminates. Specifically we calculate the component of residual corresponding to the current mesh point at the end of each Jacobi iteration kernel call and set a global flag if its absolute value is above a given threshold. For a typical spatial discretization $h_x = h_y = h_z = 0.5\xi_0$ and temporal discretization $h_t = 0.1 t_0$, this method converges in about 4--6 iterations to an accuracy of $\Vert {\bf r}^{(l)} \Vert_{\max}^2 = 10^{-6}$ for a steady state configuration of flowing vortices.

To calculate the electric field, a third kernel integrates the ODE~(\ref{eq:K_ODE}) for $K$, again using one thread per grid point of the mesh. Calculating $J_{\rm av}$ requires a summation over all the mesh points, so a fourth kernel is used to perform a reduction over the output of all the threads (using a recommended parallel reduction kernel from the CUDA Toolkit). The final steps of the reduction are performed on the host, which returns a value for~$K$ to the GPU.

Analogous to the TDGL solution, but using an additional pseudo time relaxation $\tau$, a fifth and sixth kernel are used to solve the Poisson equation for the scalar potential. Due to the over-relaxation method, matrix ${\cal N}$ of Eq.~(\ref{eq:poissonlinear}) is diagonally dominant guaranteeing convergence of the Jacobi iterations.

Virtually all the calculations in a time step are performed on the GPU device and only a small set of floats is transferred to and from the GPU to the host,  aside from checkpointing or saving the state of the system for post-processing.

The algorithm is implemented straight forwardly on GPUs using CUDA and as a single-thread version on CPUs (not using GPU emulation). Optimizations for a specific GPU architectures were not performed. Comparing parts of the algorithm to highly optimized stencil calculations on specific GPU architectures suggests that further optimization of the kernels would improve performance by at most a factor of two.  However, implementing these optimizations is beyond the scope of the present work, as is deploying optimized implementations for other parallel architectures. Rather, in the following sections, we provide examples of the size, scale, and runtime requirements of a set of simulations capable of addressing meaningful scientific questions that our formulation and implementation has enabled. We also investigate the performance and scaling of the different parts of the algorithm in response to changing the parameters of the simulation such as discretization, accuracy, and system size.

\subsection{Example simulations} \label{sec:results}

\begin{figure*}[tb]
    \begin{center}
        \subfloat{\includegraphics[width=13.8cm]{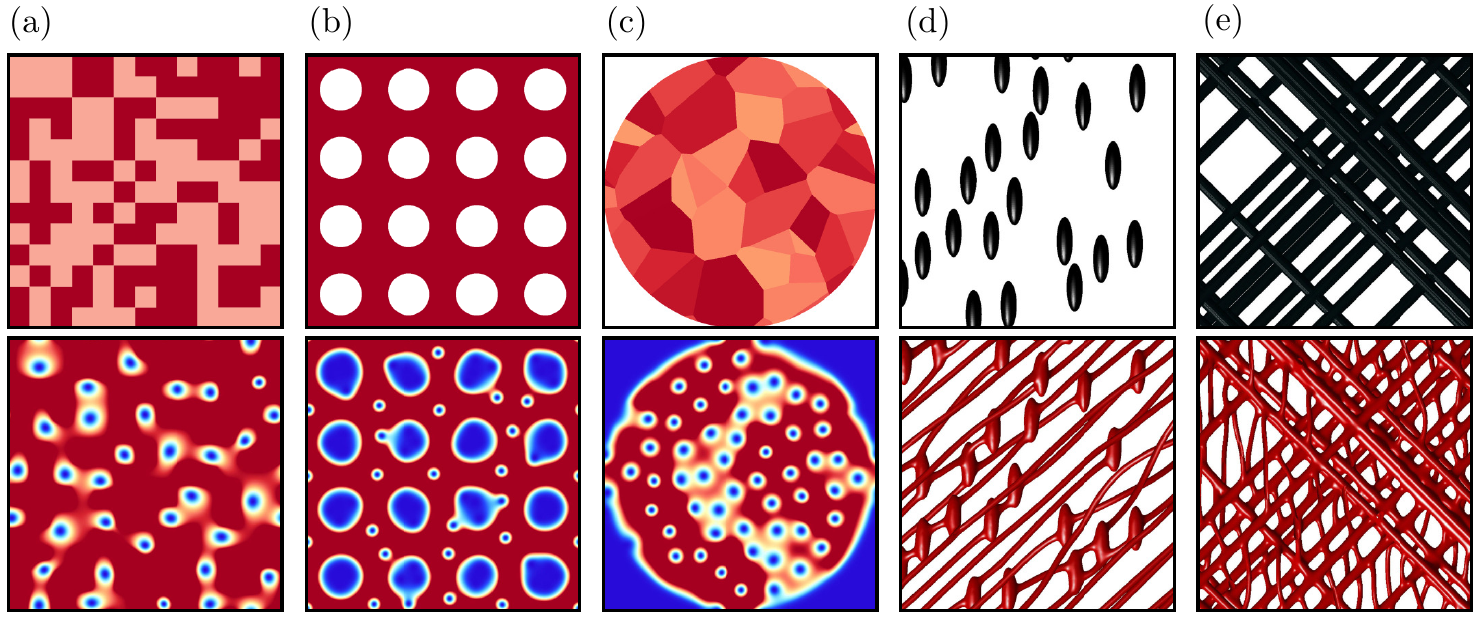}\label{fig:patterns_a}}
        \subfloat{\label{fig:patterns_b}}
        \subfloat{\label{fig:patterns_c}}
        \subfloat{\label{fig:patterns_d}}
        \subfloat{\label{fig:patterns_e}}
    \end{center} \vspace{-4mm}
    \caption{
        Patterns $\epsilon({\bf r})$ (upper row) and corresponding order parameter
        $|\psi({\bf r})|^2$ as density plot in 2D or isosurface plot in 3D (lower row).
        (a)~2D random checkerboard of two different~$\Tc$ values in perpendicular magnetic field.
        (b)~2D square lattice of circular inclusions.
        (c)~2D Voronoi/polycrystalline pattern in a circular domain.
        (d)~randomly distributed elliptical inclusions in $45^\circ$ tilted magnetic field in 3D.
        (e)~$\pm 45^\circ$ oriented columnar defects in vertical magnetic field in 3D.
    }
    \label{fig:patterns}
\end{figure*}

\begin{figure*}[tb]
    \begin{center}
        \subfloat{\includegraphics[width=11.2cm]{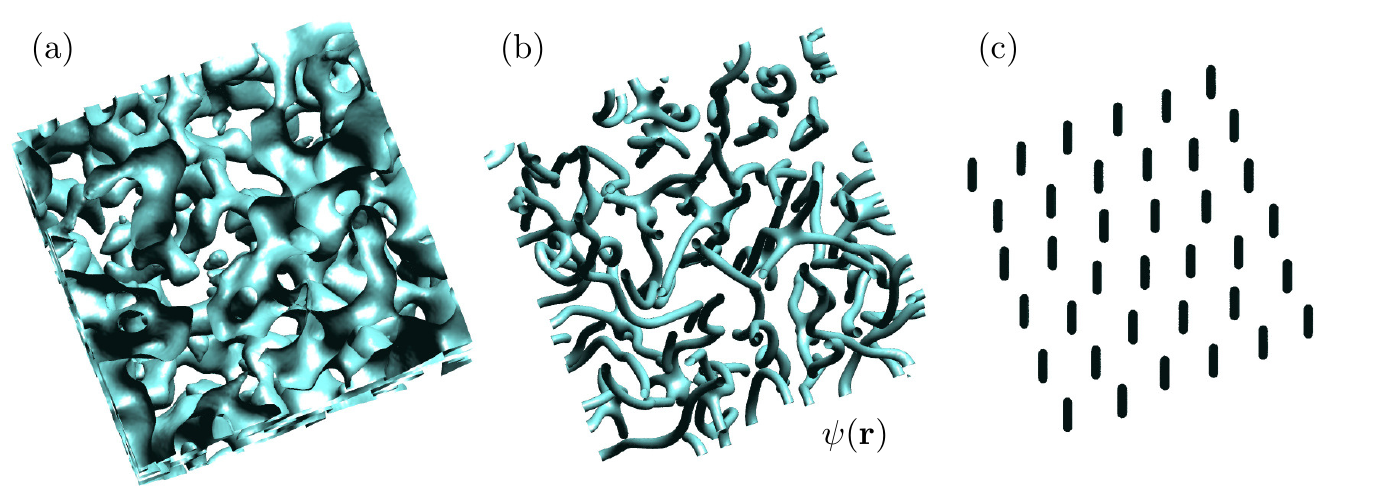}\label{fig:relaxation_a}}
        \subfloat{\label{fig:relaxation_b}}
        \subfloat{\label{fig:relaxation_c}}
    \end{center} \vspace{-6mm}
    \caption{
        Isosurfaces of the absolute value of the order parameter 
        $|\psi({\bf r})|^2$ in a cubic system of size $(128\xi_0)^3$ 
        in a presence of magnetic field. The order parameter relaxes 
        from random distribution to Abrikosov lattice.
        (a)~Nearly random spatial distribution of the order 
        parameter after $N_t = 60$ time steps.
        (b)~Beginning of the relaxation, $N_t = 150$. 
        (c)~Steady state (vortex lattice), $N_t = 10^4$.
    }
    \label{fig:relaxation}
\end{figure*}

\begin{figure}[tb]
    \begin{center}
        \includegraphics[width=8.6cm]{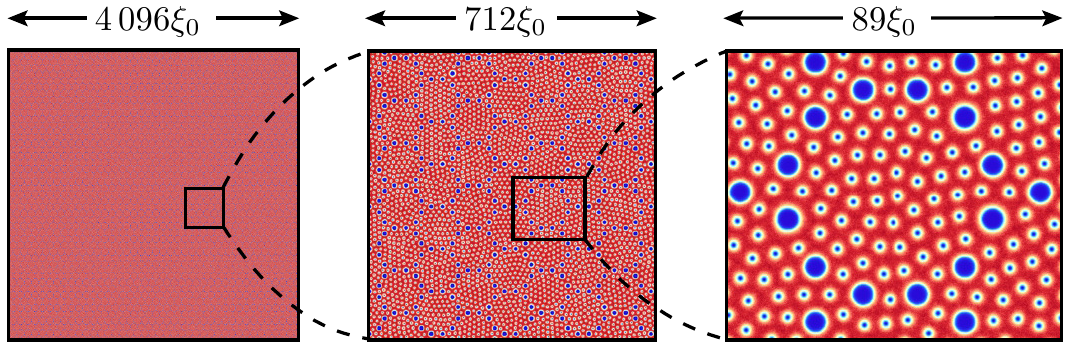}
    \end{center} \vspace{-4mm}
    \caption{
        Plot of $|\psi({\bf r})|^2$ for simulation of mesh size
        $N_x \times N_y = 8\,192 \times 8\,192$ of a large 2D kagome 
        ice lattice (large blue circles). The shown configuration has more
        than 250\,000 vortices (small blue circles).
    }
    \label{fig:scalability}
\end{figure}

Using the pattern generator, we demonstrate the variety of systems that can be modeled. Figure~\ref{fig:patterns} shows several examples of 2D and 3D inhomogeneous systems, where the inhomogeneity is modeled via a spatial dependence of the critical temperature. In the upper panel, different materials, which correspond to different values of $\Tc$, are shown as shades of red, while non-superconducting regions are white in the 2D cases, Fig.~\subref{fig:patterns_a}--\subref{fig:patterns_c}, or black in the 3D cases, Fig.~\subref{fig:patterns_d} and \subref{fig:patterns_e}. The lower panels of Fig.~\ref{fig:patterns} show snapshots of the same systems in steady state. The bottom panel of each 2D system is colored by the absolute value of the order parameter using the same color scheme as Fig.~\ref{fig:scalability}. For the 3D figures, Fig.~\subref{fig:patterns_d} and \subref{fig:patterns_e}, isosurfaces of $|\tpsi({\bf r})|^2$ are shown in red.

Figure~\subref{fig:patterns_a} shows a system initialized with a random map of two different~$\Tc$ values ($\epsilon = 1$ in red and $\epsilon = 0.5$ in light red) with equal probability. In the bottom frame, the expected low-energy Abrikosov lattice of vortices has been disrupted and, instead, the vortices predominantly exist in the lower $\Tc$ domains.

Figure~\subref{fig:patterns_b} shows a square lattice of circular non-superconducting inclusions (white with $\epsilon = 0$ inside). These inclusion act as pinning centers. In the lower frame, the lattice of large blue circles reflect the suppression of the superconducting field inside the inclusions. Observing the system over time, vortices (smaller blue circles) move between the pinning sites. Using this method of visualizing the $\tpsi({\bf r})$ field, vortices pinned inside of the pinning centers cannot be observed.

Figure~\subref{fig:patterns_c} shows a standard Voronoi tessellation pattern modeling heterogenous superconducting crystallites of different sizes inside a circular integration domain. In the panel below, the band of weak superconductivity through the crystallites with smaller~$\Tc$ (lighter color in top panel) is apparent. Also, it is clear that the diameter of vortices in different crystallites depends on the~$\Tc$ of the crystallite.

In Fig.~\subref{fig:patterns_d} ellipsoidal inclusions have been randomly placed in 3D superconducting media where a magnetic field is applied at $45^{\circ}$. The isosurface of the order parameter in the lower panel shows a pinned vortex configuration. While vortices are aligned with the magnetic field in the bulk superconducting media, the aspect ratio of the inclusions creates an anisotropic pinning behavior. Indeed, simulations of this system show that the critical current depends on the angle of the applied magnetic field relative to the orientation of the elliptical inclusions.

Figure~\subref{fig:patterns_e} shows columnar defects randomly placed and oriented at angles of $\pm 45^{\circ}$ in a 3D superconducting media where a magnetic field is applied in the $z$ direction. In the lower panel, one can observe how the $z$-axis aligned vortices bend and align through the angled defects.

In Fig.~\ref{fig:relaxation} we show the relaxation of a 3D homogeneous superconducting system penetrated by a magnetic field. Specifically shown is the Isosurfaces (cyan) of the absolute value of the order parameter $|\psi({\bf r})|^2$. The $|\psi({\bf r})|$ field is randomly initialized. Next a disordered state with complex vortex structures [Figs.~\subref{fig:relaxation_a} and \subref{fig:relaxation_b}] forms. Finally the system relaxes to a 3D Abrikosov lattice [Fig.~\subref{fig:relaxation_c}].

Figure~\ref{fig:scalability} shows a large inhomogeneous 2D system relaxed to a steady state configuration. The system of equations were solved on a mesh of $N_x \times N_y = 8\,192^2$ grid points, or a system of physical size $(4\,096\xi_0)^2$, where~$\xi_0$ is the coherence length at zero temperature. This system size is roughly the size of a typical experimental setup for a 2D superconducting film. Using the pattern generator, a kagome ice perforated lattice of inclusions was applied to the system. The three frames of Fig.~\ref{fig:scalability} shows, at different length scales, the value of $|\psi({\bf r})|^2$ on a red-to-white-to-blue color scale, where red is the maximum value, that is, where the material is the most superconducting, and blue is the minimum value, that is, where the superconducting field is completely suppressed. The kagome lattice of inclusions is apparent as the large blue circles visible in the final frame of Fig.~\ref{fig:scalability}. The small blue circles are vortices. The system shown in Fig.~\ref{fig:scalability} holds more than 250\,000 vortices.

\subsection{Performance and scalability}
\label{sec:performance}

Here we analyze the scalability and relative performance of parts of the algorithm using our implementation on GPUs. We concentrate on application relevant examples, namely, a large 2D and 3D simulation of a vortex lattice ``free flowing'' in response to an external current. 

For both the 2D and 3D system, we simulate a disorder-free system with periodic boundary conditions with a magnetic field applied in $z$ direction and an external currents in $x$ direction.  The vortices in the superconductor form a Abrikosov vortex lattice which moves in $y$ direction.  The 2D system is modeled by $N = 8\,192^2 = 67\,108\,864$ grid points.  Simulating $N_t = 10^4$ time steps required approximately 21 and 37 minutes in single and double precision, respectively, on a NVIDIA K20Xm GPU. The accuracy of the solution at each time step is $10^{-6}$ with the convergence criterion described above. In comparison, a 3D system of free flowing vortices with $N = 406^3 = 66\,923\,416$ grid points, simulated for the same number of time steps on the same hardware required 33 and 55 minutes in single and double precision, respectively. For this relatively short simulation time, the 3D simulation is $\sim 1.5$ times slower than the 2D simulation because of the calculation of the third component of the Laplacian and the slower relaxation to the steady state in 3D.  The choice of boundary condition has no measurable influence on the runtime.

\begin{figure*}[tb]
    \begin{center}
        \subfloat{\includegraphics[width=6.5cm]{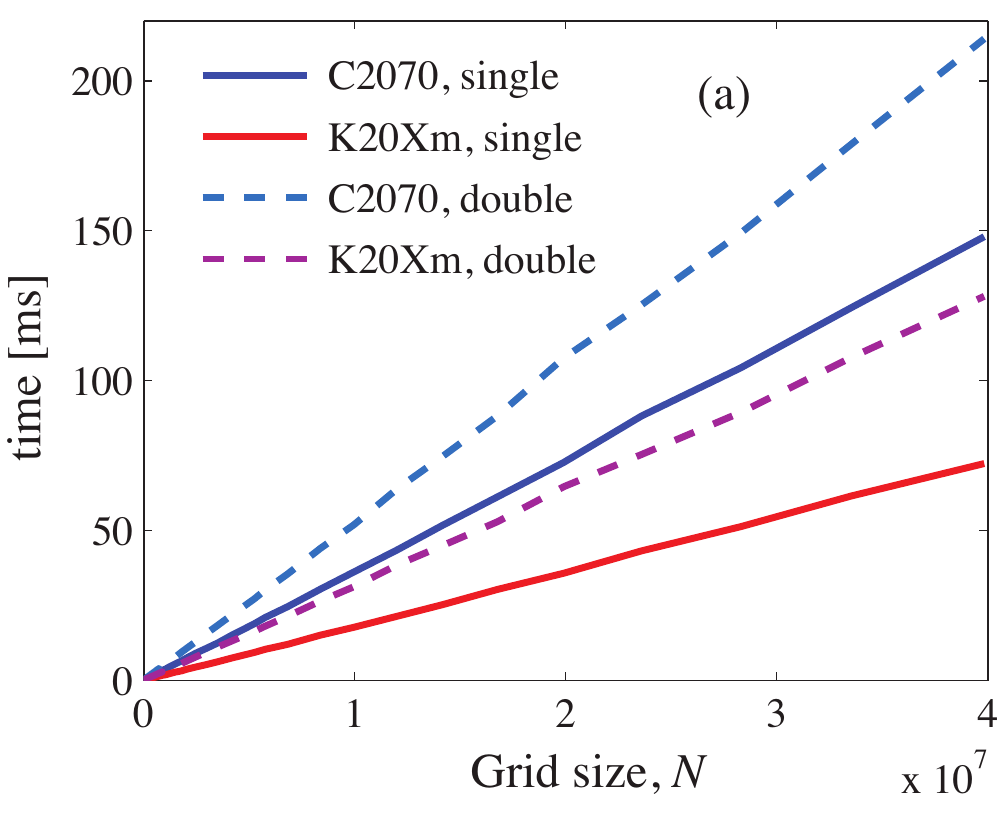}\label{fig:runtime}} \hspace{0.3cm}
        \subfloat{\includegraphics[width=6.5cm]{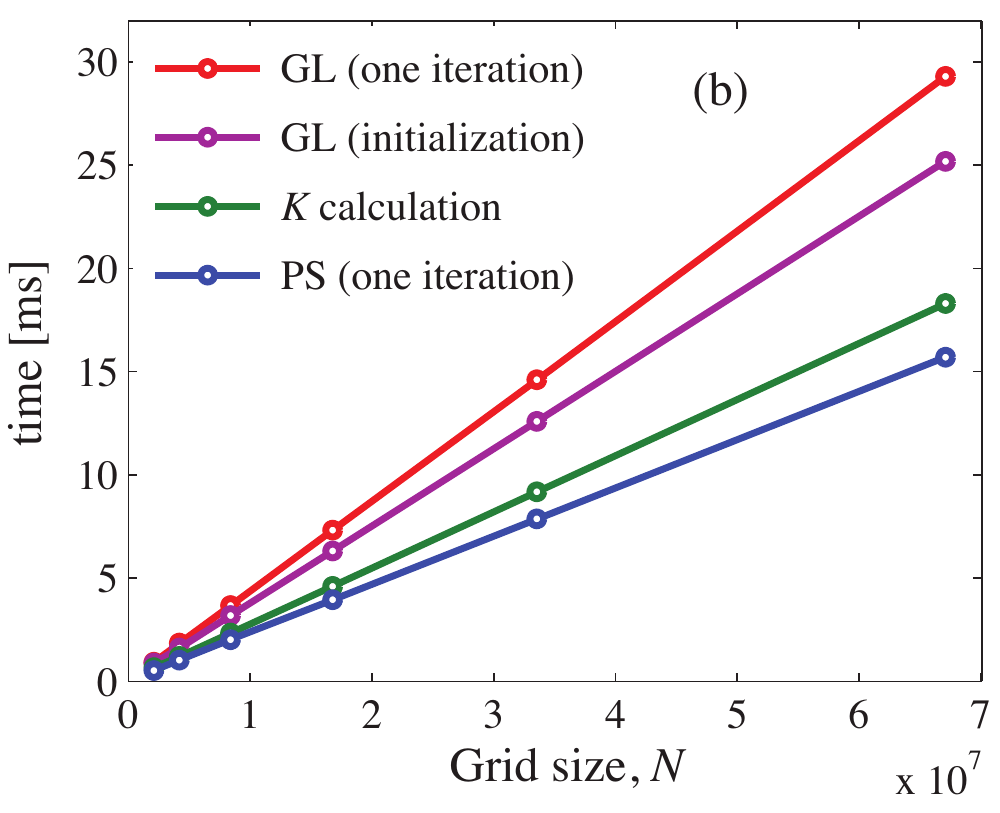}\label{fig:profile}}
    \end{center} \vspace{-6mm}
    \caption{
        (a)~Simulation time in milliseconds for one time step 
        using single or double precision for Tesla C2070/K20Xm GPUs.
        Simulation times are for a 2D $N_x \times N_y$
        homogeneous [$\epsilon({\bf r}) = 1$] system (superconducting film).  
        The grid size is increased by varying~$N_y$ while $N_x = 1024$. 
        The external current is applied in $x$ direction and the external 
        magnetic field perpendicular to the current in $z$ direction.
        (b)~Profile of a single time step as a function of system size 
        (single precision 3D simulation using a Tesla K20Xm).
    }
    \label{fig:performance_comparison}
\end{figure*}

To investigate the scaling of our implementation, we simulate the described 2D and 3D systems while varying the system size.  Figure~\subref{fig:runtime} shows the runtime for integrating a single time step for various system sizes on NVIDIA GPUs. Both single and double precision versions are tested on Tesla C2050/C2070 (1.15\,GHz, 448 cores) and Tesla K20Xm (732\,MHz, 2\,688 cores) GPUs.  The time to calculate a single time step is averaged over $N_t = 10^4$ time steps after the system has reached steady state. The time scales linearly with the number of mesh points for $N > 10^6$ mesh points, above which the GPU cores are fully utilized. A 3D systems scales similarly with increasing~$N$ (not shown), but requires about 30\% longer per time step integration, which one would expect due to the third component of the Laplacian. The number of Jacobi iterations required to converge, given the same convergence criterion, is the same for the two 2D and 3D steady state simulations.

In Figure~\subref{fig:profile}, we show how the different components of the time step integration contribute to the total time as a function of system size for a 3D simulation simulated on a Tesla K20Xm. These components, which correspond to the GPU kernels executed to complete a single time steps, are the initialization of the matrix for the GL equation~(\ref{eq:GL_linearized}), one Jacobi iteration for solving the GL equation, the calculation of the function $K$, Eq.~(\ref{eq:K_ODE}), and one iteration of the Poisson solver, Eq.~(\ref{eq:Poisson}). Each kernel assigns a single thread to each grid point and calculates a value for that mesh point value using only the value of a small number of neighboring grid cells, except for the calculation of $K$, which requires a summation over all the grid points.  This summation is performed using a reduction kernel recommended by NVIDIA CUDA SDK.  Figure~\subref{fig:profile} confirms that, when the problem is sufficiently large, such that GPU is fully utilized, all of the calculation components scale linearly with system size.  Since, in a typical dynamic steady state, six to seven Jacobi iterations are required per time step, the calculation of the Jacobi iterations tends to dominate the overall time.  In comparison, the Poisson solver converges very quickly and typically accounts for about 20\% of the time step.  The summation over all grid points in the calculation of~$K$, and the initialization of the GL equations, thus, account for only a small part of a typical time step. 

\begin{figure*}[tb]
    \begin{center}
        \includegraphics[width=4.1cm]{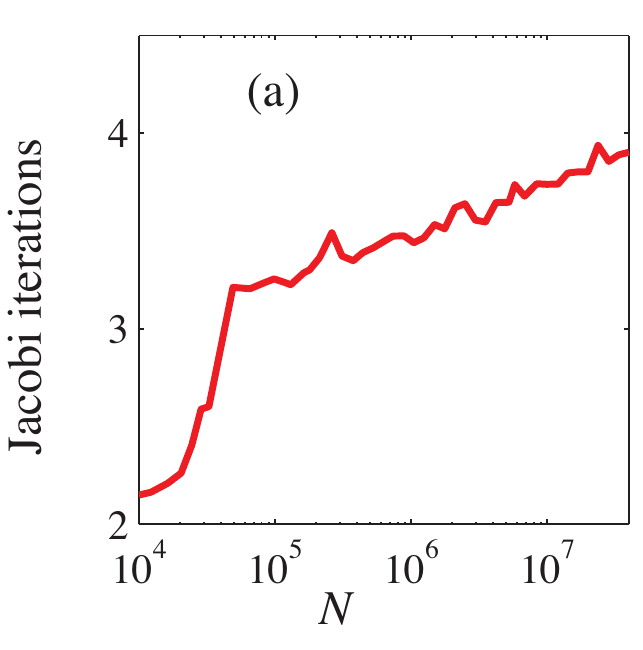} \hspace{3mm}
        \includegraphics[width=4.1cm]{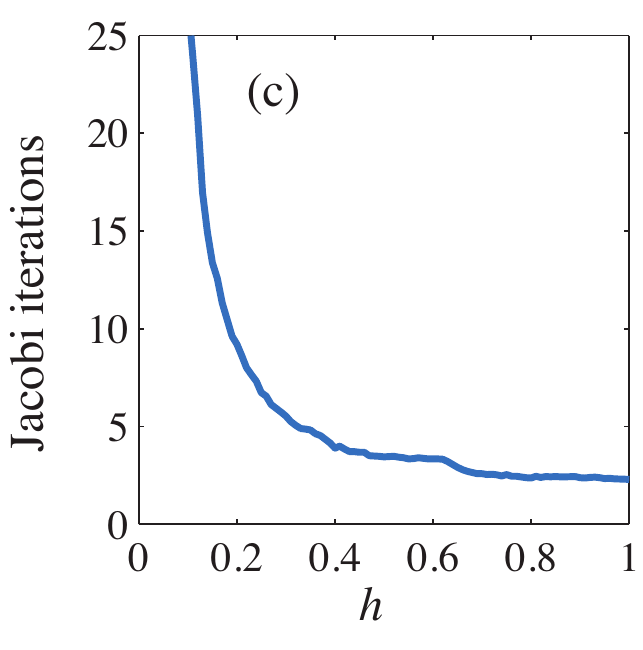} \hspace{3mm}
        \includegraphics[width=4.1cm]{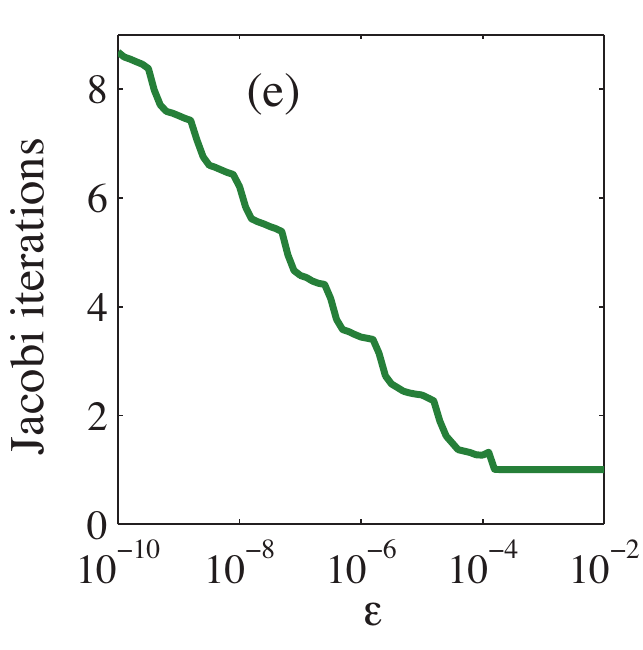} \\ \vspace{-4mm}
        \includegraphics[width=4.1cm]{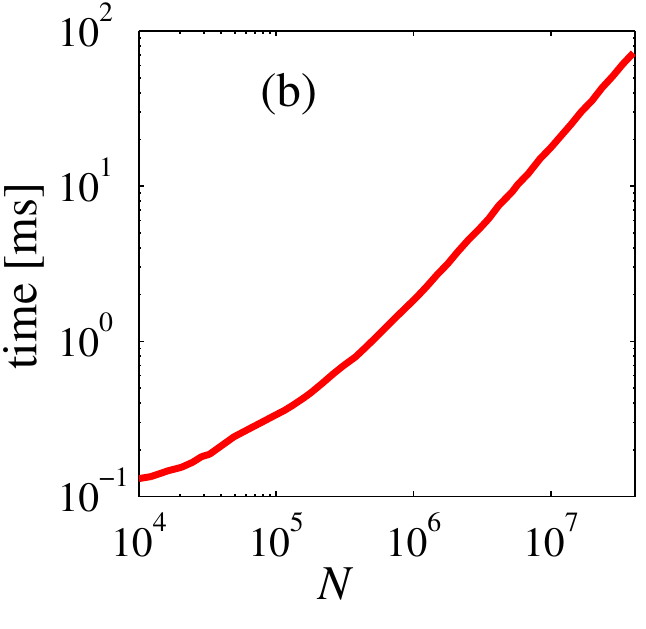} \hspace{3mm}
        \includegraphics[width=4.1cm]{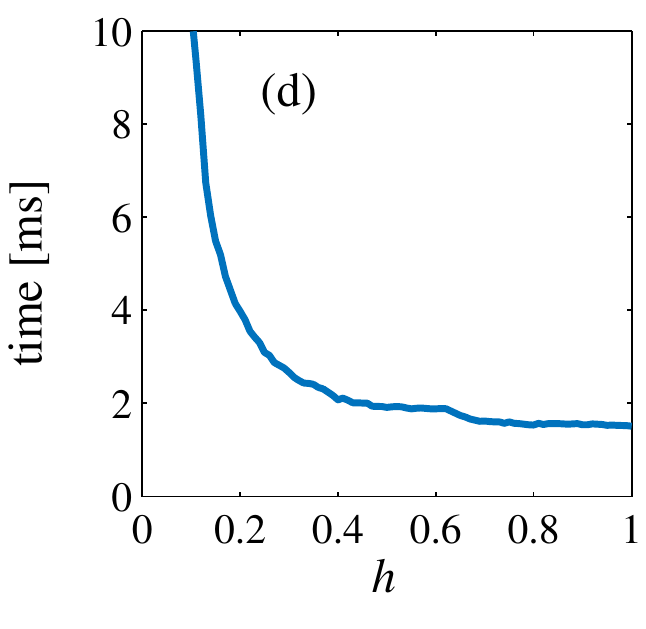} \hspace{3mm}
        \includegraphics[width=4.1cm]{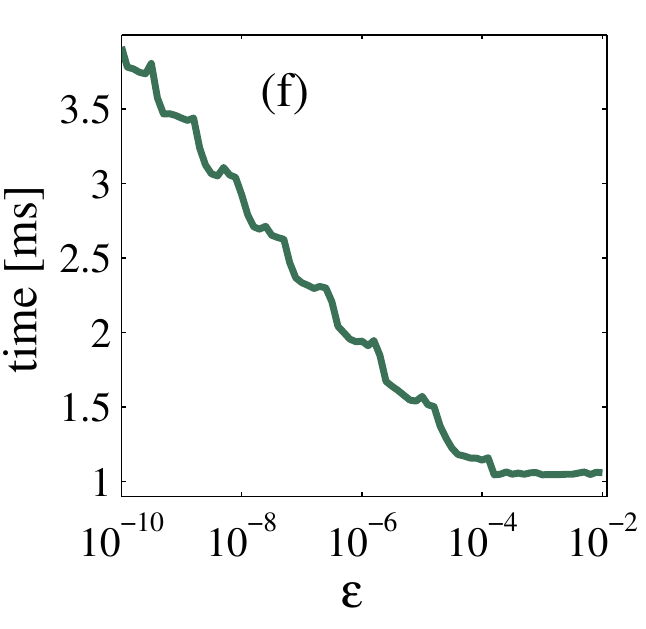}
        \subfloat{\label{fig:iterations_Nx}}
        \subfloat{\label{fig:runtime_Nx}}
        \subfloat{\label{fig:iterations_h}}
        \subfloat{\label{fig:runtime_h}}
        \subfloat{\label{fig:iterations_eps}}
        \subfloat{\label{fig:runtime_eps}}
    \end{center} \vspace{-7mm}
    \caption{
        Dependence of the number of Jacobi iterations 
        and average clock time for one time step on
        (a)--(b)~grid size~$N$,
        (c)--(d)~grid discretization~$h$, and  
        (e)--(f)~convergence accuracy~$\varepsilon$. 
    }
    \label{fig:performance_parameters}
\end{figure*}

Finally, in Fig.~\ref{fig:performance_parameters} we investigated the dependence of the number of Jacobi iterations needed to solve Eq.~(\ref{eq:GL_linearized}) on the system size, mesh discretization, and accuracy $\varepsilon$  in 2D, and the resultant impact on the simulation clock time.  Simulations were performed on a Tesla K20Xm with single precision for the 2D free-flow sample with $h = h_x = h_y$.  If not otherwise stated, $N_x = N_y = 1024$, $h = 0.5\xi_0$, and  $\varepsilon$ is set to $10^{-6}$.   The upper row of Fig.~\ref{fig:performance_parameters} shows the number of Jacobi iterations needed and the lower row the clock time for an average time step.    Since the number of Jacobi iterations required is the dominant component of the time to compute a single time step, the two curves are highly correlated.

Figures~\subref{fig:iterations_Nx} and \subref{fig:runtime_Nx} show the dependence on system size. The number of Jacobi iterations depends slightly on system size, but only logarithmically.  As seen previously in Fig.~\subref{fig:runtime}, the clock time depends linearly on the system size for sufficiently large systems.  

Figures~\subref{fig:iterations_h} and \subref{fig:runtime_h} show the dependence on the mesh discretization~$h$. As $h$ becomes smaller the number of Jacobi iterations, and consequently the clock time, diverge. This implies that increasing the accuracy of solutions by refining the mesh comes with a significant additional cost. We note, however, that the GL equations have physical meaning at the length scale of the coherence length $\xi_0$. Therefore, a discretization of $h \ll 0.5\xi_0$ will not contribute physically meaningful details, nor significantly change the coarser scale solution. 

Finally, Figs.~\subref{fig:iterations_eps} and \subref{fig:runtime_eps} show the dependence on the required accuracy~$\varepsilon$ of the Jacobi iteration. The calculation converges exponentially fast.  This suggests that higher precision results can be generated for just a small additional computational cost. However, in practice, the addition of thermal Langevin noise to the system means that increasing the required accuracy of the solution does not qualitatively change the results. Therefore, an $\varepsilon = 10^{-6}$ is a reasonable choice for most applications.

\section{Conclusions} \label{sec:conclusion}

In this paper we present a method to solve the TDGL equations for {type-II} superconductors for large-scale application. We pose the solver in the context of modeling superconductors in externally applied magnetic fields and currents, an important framework for studying vortex dynamics in mesoscopically large systems. The TDGL model is an especially attractive computational framework for studying the collective dynamics of vortices as the model reproduces the long range interactions and mechanisms of vortex cutting and recombination. We show how arbitrary pinning configurations can be implemented, enabling the investigation of critical currents in superconductors under a variety of conditions.

The formalism and solver are described in detail. Time integration schemes are chosen to create numerically stable solutions. We show that system sizes as large as $\sim (6\,000\xi_0)^2$ for 2D superconducting films or $\sim (270\xi_0)^3$ for 3D cubic superconductors can be implemented on a single GPU. While the zero-temperature coherence length, $\xi_0$ in type-I superconductors can be large (1.6\,$\mu$m in aluminum), it is typically 10--100\,nm in conventional {type-II} superconductors, and 2--5\,nm in high-temperature materials. Overall, this means that, even on a single GPU, experimentally relevant system sizes can be simulated. That is, our formulation makes studying problem approaching an experimental size and time-scale computationally tractable.

We have implemented a prototype of our algorithm on a GPU. This implementation is already sufficient to allow for large-scale problems to be studied, on larger scale than any other work in the field we are currently aware of, in reasonable amounts of computational time, e.g. hours rather than days or weeks. Additional work is required to optimize the implementation for different massively parallel computational environments (e.g. single GPUs, multiple GPUs, a cluster of multi-threaded CPU nodes connected by a fast network). However, we believe that this formulation is highly amenable to parallelization in such environments. Also, while here we present a formulation and discretization of our domain on a regular mesh, further generalization to a discretization on unstructured meshes using the presented methodology should be possible as well.

Aside from investigating {type-II} superconductors, the presented methodology can be readily adapted for a much broader range of systems, which can be described by differential equations similar to TDGL equations, ranging from cold-atom simulations~\cite{Glatz:2011}, to fluid dynamics applications~\cite{Newell:1993,Cross:1993}, to field theories that describe liquid crystals and superfluids~\cite{Pismen:1999}. Thus our formulations of the solver is sufficiently general to be applicable to a wide-range of problems in fundamental science and important to energy related technologies.

We are delighted to thank I. S. Aranson for useful discussions. The work was supported by the Scientific Discovery through Advanced Computing (SciDAC) program funded by U.S. Department of Energy, Office of Science, Advanced Scientific Computing Research and Basic Energy Science. C.L.P. was funded by the Office of the Director through the Named Postdoctoral Fellowship Program (Aneesur Rahman Postdoctoral Fellowship), Argonne National Laboratory.

\bibliographystyle{apsrev4-1}
\bibliography{GLGPU}

\end{document}